\documentclass[sigconf,balance=false]{acmart}
\PassOptionsToPackage{table,dvipsnames}{xcolor}
\usepackage{popets}

\setcopyright{popets}
\copyrightyear{2025}

\acmYear{2025}
\acmVolume{YYYY}
\acmNumber{X}
\acmDOI{XXXXXXX.XXXXXXX}
\acmISBN{}
\acmConference{Proceedings on Privacy Enhancing Technologies}
\usepackage[T1]{fontenc}
\usepackage{hyperref}
\usepackage{graphicx}
\usepackage{caption}
\usepackage{subcaption}
\usepackage{wrapfig}
\usepackage{tabularx}
\usepackage{adjustbox}
\usepackage{listings}
\usepackage{color} 
\usepackage{enumitem}
\usepackage{pifont}
\newcommand{\cmark}{\ding{51}}

\usepackage{multirow}
\usepackage{longtable}
\usepackage{amsthm}
\usepackage{balance}
\usepackage{multicol}
\usepackage{url}
\usepackage{colortbl}

\usepackage{array}
\hypersetup{
    colorlinks = false,
}

\newcommand\tcfapi{\texttt{\_\_tcfapi}}
\newcommand\onetrustg{\textsf{OTAG}}
\newcommand\optanon{\texttt{OptanonConsent}}
\newcommand\tcs{{TCString}}
\newcommand\tcss{{TCStrings}}

\usepackage[htt]{hyphenat} 

\newif\ifediting

\ifediting
    \usepackage{comment}
    \usepackage[draft,textsize=footnotesize,textwidth=17mm]{todonotes}
\fi

\def\inlinedremark#1#2#3{        
  \ifediting
        \textcolor{#2}{\textbf{#1: } #3}
    \fi
    }

\def\NBcom#1{\textbf{\REMARK{Nataliia}{\textcolor{orange!100}{#1}}}}

\def\GKcom#1{\textbf{\REMARK{Gayatri}{\textcolor{magenta!100}{#1}}}}

\def\NBtext#1{\inlinedremark{Nataliia}{orange!100}{#1}}

\def\GKtext#1{\inlinedremark{Gayatri}{magenta!100}{#1}}

\def\REMARK#1#2{        
  \ifediting
    \begin{center}
    \noindent\framebox{
        \begin{minipage}{0.9\columnwidth}
            \textrm{{[#1]: #2}}
        \end{minipage}}
    \end{center}
    \fi
    }

\begin{document}


\title[Measuring Compliance of Consent Revocation on the Web]{Johnny Can't Revoke Consent Either: \\
Measuring Compliance of Consent Revocation on the Web}
  
\author{Gayatri Priyadarsini Kancherla}
\affiliation{%
\institution{IIT Gandhinagar}
\city{Gandhinagar}
\country{India}}
\email{gayatripriyadarsini@iitgn.ac.in}

\author{Nataliia Bielova}
\affiliation{
\institution{ Inria Centre at University Côte d'Azur}
\city{Sophia Antipolis}
\country{France}}
\email{nataliia.bielova@inria.fr}

\author{Cristiana Santos}
\affiliation{ 
\institution{Utrecht University}
\city{Utrecht}
\country{Netherlands}}
\email{c.teixeirasantos@uu.nl}

\author{Abhishek Bichhawat}
\affiliation{
\institution{IIT Gandhinagar}
\city{Gandhinagar}
\country{India}}
\email{abhishek.b@iitgn.ac.in}


%


\begin{abstract}
The EU General Data Protection Regulation (GDPR) requires websites to facilitate the right to revoke consent from Web users. Prior works have examined consent management by 
auditing that user choices are correctly stored, and comparing cookies set upon acceptance versus rejection to assess compliance. While these studies measured compliance of consent with respect to the various consent requirements, no prior work has studied consent revocation on the Web. Therefore, it is unclear how  difficult it is to revoke consent  on the websites' interfaces, and whether the revoked consent is properly stored and communicated behind the user interface. 


Our work aims to fill this gap by measuring compliance of consent revocation on the Web  on Tranco's top-200 websites.
We found that 19.87\% of websites  make it difficult for users to revoke consent throughout different interfaces, 20.5\% of websites require more effort than acceptance, and  2.48\% do not provide consent revocation at all,  thus violating EU legal requirements for valid consent. 
%
57.5\% {websites do not delete the cookies after consent revocation enabling continuous illegal  processing of users' data.} 

Further, we analyzed 281 websites implementing the IAB Europe Transparency \& Consent Framework, and found 22 websites that store a positive consent despite  user's revocation. 
%
Surprisingly, we found that on 101 websites, third parties that have received consent upon user's acceptance, are not informed of revocation, leading to the illegal processing of users' data by such third parties according to EU laws. 
%
Our findings emphasize the need for improved legal compliance of consent revocation, and proper, consistent, and uniform implementation of revocation communication 
to third-parties.

%

\end{abstract}

\maketitle




\NBcom{Gayatri, please see my comments with NBcom and NBtext through the document and also in the figures of the appendix.}

\section{Introduction}



In recent years, compliance with privacy and data protection laws on the Web has gained a lot of attention, both from  the research community and regulators. The EU Data Protection framework, consisting of the General Data Protection Regulation (GDPR)~\cite{gdpr} and ePrivacy Directive (ePD)~\cite{ePD-09} sets requirements for a legally-valid consent when tracking technologies are deployed on a website or app. Such consent is usually implemented via consent banners and must 
satisfy seven requirements for validity: it must be prior to any data collection, freely given, specific, informed, unambiguous, readable and accessible, and revocable~\cite[Art. 4(11), 7]{gdpr},~\cite[para 17]{Sant_etal_20_TechReg}.
While numerous studies have evaluated compliance of consent banners with the various consent requirements, such as the presence and visibility of the rejection button ~\cite{do_cookie_banners_respect,automating_gdpr_violation_detection,dark_patterns_after_gdpr}, one aspect has so far been overlooked by the research community and the regulators: the requirement of \emph{revocable consent}, also known as consent withdrawal, which entails that the user has the option to change a prior preference regarding trackers due to the reversible nature of consent decisions~\cite{wp208}. 


%
%

According to the GDPR,  users have the \textit{right to revoke consent at any time}~\cite[Art. 7(3), Rec. 42]{gdpr}. Accordingly, websites must \textit{facilitate} the exercise of this right~\cite[Art. 12(2)]{gdpr} by providing an option to  revoke consent. 
Upon revoking consent, websites must subsequently comply with an additional obligation to \textit{delete} data previously processed on the basis of that consent, and without undue delay even if the user did not explicitly exercise their right to request data deletion~\cite[Art. 17(1)(b)] {gdpr}.
This paper focuses exclusively on the right to revoke consent.
Websites that do not allow consent revocation are deemed to be processing data illegally, and run the risk of being fined for not complying with the legal requirements~\cite{noyb-2024-darkpatterns,DanishAgency-revokeDecision,CNIL-revokeDecision, UODO-revokeDecision,Garante-revokeDecision}. 
%

However, it remains unclear \emph{whether websites provide users with compliant methods to revoke their consent,   whether revoked consent is properly stored by the websites behind the consent interface, and whether it is communicated to 
third-parties that collected users' data}.
Our work aims to fill this gap by measuring the compliance of consent revocation on the Web,  
%
%
%
addressing the research questions:
\begin{itemize} [leftmargin=10mm,itemsep=1pt]
\item [\textbf{RQ1:}] Are the revocation interfaces on websites compliant with EU data protection laws? 
\item [\textbf{RQ2:}] Do \emph{advertising and analytics} (AA) cookies that require consent get deleted 
once consent is revoked?
\item [\textbf{RQ3:}] Is consent stored correctly in the browser, and is it consistent across browser storage and APIs implemented by {Consent Management Platforms} (CMPs)?
\item [\textbf{RQ4:}] 
Are all third parties that are initially informed of ``acceptance'' also notified when the consent is revoked? 

\end{itemize}


\noindent To address these questions, 
 we have set up a team of several computer scientists and a legal expert to
 \emph{measure compliance of consent revocation on the Web}, making the following contributions:
\begin{enumerate}[itemsep=1pt]
\item We provide the first 
in-depth legal analysis of the GDPR, ePD and 
other legal guidelines for \emph{consent revocation} interface, storage and communication, 
and establish six operational legal requirements (\S\ref{sec:legal_bg}).
\item We provide a method to audit compliance of consent revocation interfaces (\S\ref{sec:meth-data_collection}), and apply our analysis on Tranco's top-200 websites 
identifying multiple instances of non-compliance with the legal requirements (\S\ref{sec:interface}).
\item We study the effect of revoking advertising and analytics (AA) cookies, finding that on the majority of websites, revocation does not decrease the number of AA cookies, thus violating the GDPR (\S\ref{sec:cookies}).
\item We propose a methodology (\S\ref{sec:meth-rq34}) to evaluate revocation on websites that implement  \emph{IAB Europe Transparency and Consent Framework} (TCF) and the most popular Consent Management Platform, OneTrust. 
We 
 analyse 281 such websites and evaluate the storage of consent behind the consent banner interface (\S\ref{sec:consistency}) and the communication of revocation (\S\ref{sec:inform-consent-revocation}), thereby detecting multiple non-compliant practices. 


\item Finally, we give recommendations to further improve consent revocation compliance  (\S\ref{sec:disc}).
\end{enumerate}

\begin{table}
\centering
\begin{tabular}{|p{6.4cm}|p{1.45cm}|}
\hline
\textbf{Violations within the interface and cookies in the top-200 websites} 
& \textbf{Prevalence}
\\ \hline
Revocation via a different interface/medium (\S\ref{sec:interface})
& \textbf{20.25\%} (32/158) 
\\ \hline
Two or more steps to find revocation interface vs zero steps to accept (\S\ref{sec:interface})
& \textbf{20.8\%} (33/158) 
\\ \hline
Data processing based on AA cookies is not stopped upon user’s consent revocation(\S\ref{sec:cookies})
& \textbf{57.5\%} (69/120)
\\ \hline \hline
\textbf{Violations beyond the interface in 281 websites with CMPs } 
& \textbf{Prevalence}
\\ \hline
Positive consent after revocation (TCF-based consent string) (\S\ref{sec:consistency})
& \textbf{16.17\%} (22/136) 
\\ \hline
Positive consent after revocation (OneTrust-specific consent string) (\S\ref{sec:consistency})
& \textbf{14.47\%} (22/152) 
\\ \hline
Third-parties informed of consent acceptance via HTTP requests, but not informed of revocation (\S\ref{sec:inform-consent-revocation})
& \textbf{74.2\%} (101/136)
\\ \hline
\end{tabular}
  \caption{Summary of most prominent potential violations of consent revocation implementation. 
  In brackets, we mark the number of websites with violations and the total number of websites analysed for each case.}
  \label{tab:global-results}  
\end{table}

Table~\ref{tab:global-results} summarizes 
the most prevalent (more than 10\%) potential violations of the GDPR and ePD that we detected 
while analysing both the consent revocation interface, cookies, and the consent storage and sharing with third-parties. 
Among the top-200 websites,  
we find that 22.7\% (36) websites are compliant with GDPR and ePD within their revocation interfaces and management of cookies (see remaining potential violations in rows 1-3), and the compliance is more common in websites that employ CMPs. 
%
We, therefore, further investigate websites with CMPs for possible non-compliance. 
Of the 281 with CMPs, 
251 provided revocation options -- among them, we further check if 
%
a positive consent is stored even after revocation and if the consent modification is communicated to third-parties that were informed of acceptance. Overall, we find that 52.6\% out of 251 websites providing revocation do not have a possible violation beyond the interface, while the remaining websites either had positive consent after revocation or communicated an initial positive consent to third-parties but did not communicate negative consent after revocation  (rows 4-6).

\section{Background and Related Works}



In this section, we provide background on different techniques for consent management and discuss prior works on compliance in the context of consent banners, opt-out, and revocation. \\

\noindent \textbf{Consent Interfaces and Compliance.}
\label{sec:consent_interfaces}
%
%
%
Consent banners have become a common method for obtaining consent 
for online tracking.
Recent works~\cite{Sant_etal_20_TechReg, Biel-etal-24-JOLT, ndss19, imc20, acceptall, 4years} have examined consent banners following the implementation of the GDPR and the ePD while numerous others~\cite{dark_pattern_legal_req, 10.1007/978-3-031-05544-7_21, 10.1145/3491101.3519687, nordichi20, darkpattern, popets2020, asiaccs19, do_cookie_banners_respect, dark_patterns_after_gdpr, taming-cookie-monster, AW-24-popets} explored the impact of privacy regulations on consent design, and how the websites and CMPs might violate these laws, e.g., by using \emph{dark patterns}. 
%
%
Tools 
have been proposed to help users handle consent banners automatically
~\cite{i_dont_care_about_cookies, ninja_cookies, consent_o_matic, super_agent, 
automating_gdpr_violation_detection, 484c16b527a24fc9b5df872de42eb1ec, usec2018, pandit_2019}. 
Several  works~\cite{automating_gdpr_violation_detection, imc20, dark_patterns_after_gdpr, darkdialogs}  focused on automated analysis of consent banners at scale.\\ 

\noindent\textbf{IAB Europe TCF and Compliance.}
The current ``de-facto'' standard for the consent banners in the EU is the IAB Europe Transparency and Consent Framework (TCF)~\cite{TCF}. The current TCF v2.2~\cite{TCFv22} defines the \emph{pre-defined purposes} for processing personal data by Consent Management Platforms (CMPs). If also defines  the \emph{format to store the user's choice}, called  {\tcs},that includes:(a) list of enabled third-party vendors registered within the TCF; (b) list of enabled purposes among the pre-defined purposes~\cite{TCF_purposes}.

The TCF standard \emph{does not specify which of the purposes require user consent}. 
According to GDPR and ePD, only purposes that do not require user consent (and users' explicit interaction with the CMP) may be enabled in \tcs\  by default. 
Matte et al.~\cite{Matt-etal-20-APF} analysed which purposes in TCF v1.1 and v2.0 require consent according to the GDPR and ePD; however, this analysis was not extended to TCF v2.2, which we perform in this paper. 

Once a user has made their choice on the CMP interface, the CMP is required to store the corresponding {\tcs}  in the browser (however, the specific storage mechanism is not specified in TCF v2.2), or implement an API for third-parties to check the {\tcs}. 
The {\tcs} can be accessed via the {\tcfapi} function by third-parties in first-party context or via the JavaScript \texttt{postMessage} API to communicate with a special \texttt{\_\_tcfLocator} iFrame. Following TCF v1.1~\cite{tcf11-url}, {\tcs} can also be shared in the outgoing HTTP requests via URL-based methods. However, TCF v2.2 does not specify any URL parameters to be used for this. 

Prior works~\cite{do_cookie_banners_respect, Matt-etal-20-APF, Smith-WWW-24, cschecker} have focused on the analysis of 
IAB Europe TCF and its consent implementation. However, none of these works studied how CMPs manage consent revocation, which is one of the main objectives of our work.\\ 



\noindent\textbf{Consent Rejection and Revocation.} 
While multiple studies~\cite{do_cookie_banners_respect,dark_patterns_after_gdpr,automating_gdpr_violation_detection} detect and analyze the presence of a reject button on consent banners, none of the prior works evaluated the interfaces that allow \emph{revocation} of a given consent within the EU data protection framework. 
Nevertheless, since the updates of the CPRA regulation in California, multiple studies have analysed the implementation of the right to opt out of selling or sharing user data within websites. 
Tran et al.~\cite{Tran-etal-24-CHI} propose a methodology to measure compliance of opt-out links' wordings automatically found on websites subject to CCPA and CPRA~\cite{ccpa}. Liu et al.~\cite{Liu-etal-24-popets} measure the impact of opting out in the presence of advertisers and tracking mechanisms. 
Aziz et al.~\cite{AW-24-popets}  assessed the IAB CCPA Compliance Framework (analogous to IAB TCF in Europe) and detected that opt-out signals are not honored on websites. A similar latest work by Du et al.~\cite{mowchecker} detected violations regarding withdrawal interfaces in mobile applications. 


Other studies evaluated user perceptions when interacting with opt-out mechanisms. As such, 
Habib et al.~\cite{data-deletion-opt-out-choices} evaluated the usability 
of data deletion 
and opt-out options related to email communication and targeted advertising. 
Habib et al.\cite{Habib-etal-2022} evaluated how different 
instructions about revocation in the banner's text 
impact users' ability to find revocation options.
None of these works, however, 
studied whether the revocation interfaces comply with the EU laws. Additionally, no prior work has shed light on whether websites correctly store and manage revoked consent, and inform all the third-parties  that were previously informed of consent acceptance.


\section{Legal Compliance for Revocation}
\label{sec:legal_bg}

The EU Data Protection Framework provides legal principles and requirements 
that can be applied to websites and specifically to consent revocation. 
%
%
%
%
The GDPR applies to the processing of personal data~\cite{EDPB-4-07}, and
the ePrivacy Directive (ePD)~\cite{ePD-09} provides~\emph{supplementary} rules, particularly, for the use of tracking technologies~\cite{EDPB-guidelinesGDPR-ePD}. 
Whenever 
tracking data 
is stored and read from the user's device, the ePD~\cite[Art. 5(3)]{ePD-09} requires  
websites to request \textit{user's consent} 
when tracking is used 
for \emph{purposes} 
not strictly necessary for the service provided, \emph{e.g.}, advertising~\cite{EDPB-Cookie-Exemption, EDPB-PurposeLimitation}; 
purposes required for a website to operate are exempted~\cite[Recital 66]{ePD-09}.
%
The  way to assess with certainty whether consent is required is to analyze the \emph{purpose} of each tracker on a given website~\cite{EDPB-PurposeLimitation,Foua-etal-20-IWPE}. 
%



The GDPR, as well as 
guidelines from the Data Protection Authorities (DPAs) and 
from the European Data Protection Board (EU advisory board, representing all EU DPAs) 
provide  legal requirements for \emph{consent revocation}.  While  guidelines are not legally-binding, they are part of the EU framework for data protection which we apply in this work to discern when revocation methods are compliant.
A website can be held liable and fined if it fails to comply with the GDPR principles and 
requirements for valid consent, including revocation.
%
Relevant \emph{GDPR principles} 
are presented below and numbered with \textbf{P}, and 
  \emph{legal requirements for consent revocation} for websites are numbered with \textbf{LR}. 
%
%
\theoremstyle{plain}
\newtheorem{principle}{P\ignorespaces}

\begin{principle}[\textbf{Fairness}]
\label{prin1}
\normalfont
Websites must not process personal data in a  unjustifiably detrimental, discriminatory, unexpected or misleading way~\cite[Art.5(1)(a)]{gdpr},~\cite[para 17]{EDPB-05-2020consent}.
\end{principle}
%
%

\begin{principle}[\textbf{Data protection by design}]
\label{prin2}
\normalfont
Websites must implement organisational
measures and safeguards efficiently to enable the exercise of the revocation right~\cite[para 68]{EDPB2020-DPbDbDesign}, \cite[Art. 25(1)]{gdpr}; \cite[para 70]{EDPB2020-DPbDbDesign}.
Revocation options should be provided
in an objective and neutral way, avoiding any deceptive or manipulative language or design~\cite[para 16]{EDPB2024-CookiePaywalls,EDPB2022-Darkpatterns}.
If a website requires more effort
to revoke than to give consent, 
it is deploying a 
dark pattern~\cite[para 30]{EDPB2022-Darkpatterns}. 
\end{principle}


\begin{principle}[\textbf{Accountability}]
\label{prin3}
\normalfont
Websites 
 must 
  demonstrate that revocation is performed  easily and effectively~\cite[Arts. 5(2), 24(1), Rec. 74]{gdpr}.
  \end{principle}
%


%
%

\theoremstyle{plain}
\newtheorem{legal}{LR\ignorespaces}

\begin{legal}[\textbf{Right to revoke consent}]
\label{lr1}
\normalfont
Users have the \textit{right} to revoke consent  at any time~\cite[Art. 7(3), Recital 42]{gdpr}, and thus websites are obligated to \textit{facilitate} the exercise of this right~\cite[Art. 12(2)]{gdpr} by providing a consent revocation option. 
This option to revoke consent must be clearly and distinctly recognisable.
A \emph{violation} of this requirement is the absence of revocation options, which 
renders 
consent invalid~\cite[Art. 4(11)]{gdpr} and any data processed henceforth is processed illegally, without a legal basis~\cite[Art. 6(1)(a)]{gdpr}. 
\end{legal}

\begin{legal}[\textbf{Easy revocation through the same interface}]
\label{lr2}
\normalfont
Giving and revoking consent should be available through the same means/interface ~\cite[para 115]{EDPB2022-Darkpatterns} (e.g. website, app, log-on account, etc.) 
since switching to another interface would require \textit{unnecessary effort}~\cite[para 114]{EDPB-05-2020consent}. 
    %
For consent granted via a consent banner, 
DPAs disagree on which implementation should be recommended: the Dutch DPA  only recommends that revocation should be reachable within the same website~\cite{DutchDPA-revocationDoc-2024}, while 
the German DPAs insist that it is inadmissible to search a privacy policy for a revocation option~\cite{DSK-DPA-cookies-2021}.
    %
%
%
%
%
%
%
%
While the EDPB 
states a specific revocation solution cannot be imposed,  
 and a case-by-case analysis is needed~\cite{EDPB-tf-2023}, \cite[para 31-35]{EDPB2024-CookiePaywalls}, 
%
it recommends 
a permanently visible \textit{icon} 
or a \textit{link} on a standardized place~\cite[para 32]{EDPB-tf-2023}, although not referring to its location. 
%
Few DPAs propose that such options could be displayed within the privacy or cookie policies~\cite{Danish-guidelines-2021,DSK-DPA-cookies-2021}. 
%
A \emph{violation} of this requirement would be 
 proposing revocation through another 
 interface, contacting the website by email,  asking the user to delete cookies~\cite{DutchDPA-revocationDoc-2024, EDPB-05-2020consent, DSK-DPA-cookies-2021}, 
 or using opt-out options on external websites~\cite{DSK-DPA-cookies-2021}. 
 \end{legal}

\begin{legal}[\textbf{Easy revocation through the same effort and number of steps}]
\label{lr3}
\normalfont
Ease of revocation
can be measured by the time spent and the number of actions~\cite{CNIL-recoms-2020}, \cite[para 116]{EDPB2022-Darkpatterns}. 
Actions can include the number of mouse clicks, keystrokes or swipe gestures to revoke, in comparison to the number of actions required to grant consent~\cite[para 14]{EDPB-05-2020consent}; this number must be the same~\cite[p.8]{DutchDPA-revocationDoc-2024}. 
A \textit{violation} could occur when    consent obtained through one mouse-click, swipe or keystroke, but revoking takes more steps, it is more difficult to achieve or takes more time~\cite[paras 30, 114]{EDPB2022-Darkpatterns}.
\end{legal}

\begin{legal}[\textbf{Revoking requires stopping of data processing and deletion of consent-based data}]
\label{lr4}
\normalfont
%
The receiver of data \emph{must stop subsequent data processing} after revocation~\cite[para 17]{EDPB-05-2020consent}. 
%
This is especially relevant in circumstances where the controller uses a large advertising network to target individuals and track them across several websites~\cite[para 175]{EDPB2024-CookiePaywalls}. 
%
If there is no other lawful basis justifying data processing, 
data receivers \emph{must additionally delete all the data that was processed on the basis of consent}, as mandated by~\cite[Art. 17(1)(b)]{gdpr} (assuming that there is no other purpose justifying the continued retention). Such data should be deleted even in the absence of a deletion request by the user~\cite[para 119] {EDPB-05-2020consent}. 
%
A \emph{violation} could occur when 
data is still processed after revocation. 
\end{legal}

\begin{legal}[\textbf{Correct registration of consent revocation}]
\label{lr5}
\normalfont
Websites must correctly register the user consent revocation decision, and assure that the decision made by the user in the banner interface is identical to the consent that gets registered/stored by the website~\cite[Arts. 7(1), 30, Rec. 42]{gdpr},~\cite{Sant_etal_20_TechReg, Irish-guidelines-2020}(p.9). 
A \emph{violation} occurs when 
a registered consent is different from the user’s choice.
\end{legal}



\begin{legal}[\textbf{Communication of revocation to third-parties}]
\label{lr6}
\normalfont
When users revoke consent, organisations need to make sure that this is communicated to other organisations that they have shared people's personal information with~\cite{UKDPA-callCP-2024,DutchDPA-revocationDoc-2024}. 
%
%
Santos et al.~\cite[\S5.5]{Sant_etal_20_TechReg} also proposed that ``the publisher should delete the
\emph{consent cookie} and communicate the withdrawal to all
the third-parties who have previously received consent.''
Nevertheless, DPAs do not express specific requirements on informing third-parties. 
%
A \emph{violation} could occur when 
a website  does not 
implement the registration of consent revocation correctly,
or  does not communicate it to third-parties who process the data of its users~\cite[para 176]{EDPB2024-CookiePaywalls}, \cite[para 85]{CJEU-C‑129/21-2022}. 

Another important  requirement is that  revoking consent cannot be detrimental to the user.
If the consequences of revocation result in users being unable  to access the services provided by the website, and no alternative is offered, 
the right to revoke consent cannot be considered free and may be deemed detrimental~\cite{CNIL-revokeDecision}.



\end{legal}

\section{Methodology}
\label{sec:method}

Next, we describe the data collection and analysis methodology to answer our research questions. 
%
We built a semi-automated crawler based on a  Selenium-instrumented Chromium v122.0.6261.128. 
We used the crawler to collect data between March and June 2024,
within the EU. 
Table~\ref{tab:datasets} presents an overview of the three datasets we collected, which are further explained in this section.

\begin{figure*}
  \centering
  \includegraphics[width=\linewidth]{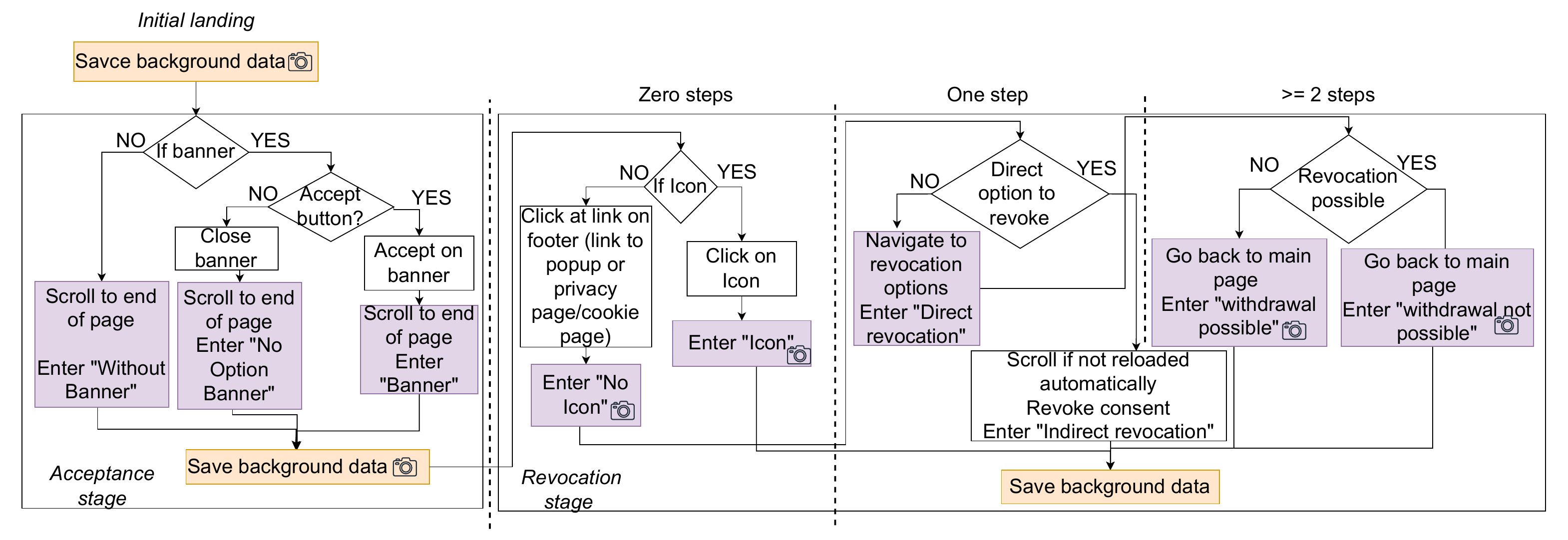}
  \caption{Data collection pipeline: to address \textbf{RQ1}, we collect screenshots and label a website as shown in violet boxes; for \textbf{RQ2} we collect cookies at each stage and also at \emph{Rejection stage} in a similar way (not shown in this figure).}
  \label{fig:data_collection}  
\end{figure*}

\subsection{Datasets for \textbf{RQ1} and \textbf{RQ2}}
\label{sec:meth-data_collection}

\begin{table}[tb]
\begin{center}
\begin{tabular}{|p{2cm}|p{3.6cm}|p{1.5cm}|}
\hline
{{\textbf{Domains}}}
& {\textbf{Collected dataset}}
&\textbf{RQs} 
\\ \hline
\multirow{2}{1.99cm}
{200 \newline({ranked 1-200})}
& \textbf{DS1}: 105 domains without \newline CMPs
  & \textbf{RQ1, RQ2} 
\\ \cline{2-3}
& \textbf{DS2}: 56 domains with \newline CMPs 
 & \textbf{RQ1, RQ2,}\newline\textbf{RQ3, RQ4} 
\\ \hline
1000 random\newline({ranked 201-5k}) 
& \textbf{DS3}: 225 domains with \newline CMPs
& \textbf{RQ3, RQ4} 
\\ \hline
\end{tabular}
\caption{Datasets used for RQs. 
}
\label{tab:datasets}
\end{center}
\end{table}

We have implemented the data collection pipeline shown in Figure~\ref{fig:data_collection} to capture the elements of the user interface that need to be interacted with in order to exercise consent revocation. At each stage of our analysis, we save background data to further analyse cookies, consent storage, and communication. Figure~\ref{fig:data_analysis} presents the details of the background data collection.

\subsubsection{Website selection}
We analyzed the revocation interface options on Tranco top-200 domains~\cite{tranco,trancolist}. 
158 domains were reachable while 42 did not display a webpage 
because they were either CDNs, blocked or returned failures. 
We detected CMPs by decoding the {\tcs} returned by {\tcfapi} following the methodology of Matte et al.~\cite{do_cookie_banners_respect}, after accepting consent. 
We detect the websites using OneTrust CMP by the presence of {\optanon} cookie, specific for this CMP. 
Of the 158 domains, 56 used TCF and/or OneTrust (labeled ``with CMP''), which we could detect. 


\subsubsection{Analyzing interfaces for consent revocation (RQ1)}
\label{sec:meth-interface}
We collected the data (including cookies, localstorage, request-response pairs and screenshots) in a semi-automated manner as shown in Figure~\ref{fig:data_collection}.Manual effort was required to browse and \textit{locate} revocation options, while all technical data (cookies/storage, screenshots, logs) were collected automatically. Automating this would require robust ML/NLP models, which may not generalize well due to the diversity in UIs and revocation paths, as discussed in results of RQ1, section \ref{sec:interface_1}.  
While we manually visited and navigated the websites, the data was collected automatically in the background. We took screenshots to record the user interfaces involved in reaching a revocation option. 
%
%
While recent works~\cite{cookie-enforcer, prisec, Bouh-etal-24-USENIX} provide automated tools to select initial consent options on websites (specifically, button and toggle HTML elements) based on user choices, our analysis shows that  
\emph{revoking consent 
may require navigating multiple links} on the page to reset the options once the initial banner is closed. Moreover, the \emph{location of the consent modification options on websites are non-standard}, 
and it is not clear whether (and where) such options are 
present. Therefore, revocation options are 
difficult to 
find and interact with automatically due to the heterogeneity of revocation implementations. 
%
%
%
%
%
%
We collected this data in four stages: 
\begin{description}[noitemsep]
    \item (1) \emph{Initial landing:} We visit each website afresh with storage cleared for all websites in the browser, and collect the screenshots showing the presence of a banner. 
    \item (2) \emph{After acceptance:} We 
    record 
    whether a banner is present 
    and if an option is provided to the user to accept consent, in which case we 
    accept all cookies. 
    \item (3) \emph{After revocation:} 
    After accepting consent, we check if an icon for modifying consent is visible on the page. If not, we check if an option is available in the footer or elsewhere on the main page. Finally, we check if the option is available on a different page reachable from the main page. For each of the options, we record the number of steps needed for revocation. 
    If no option is found, we record that revocation is not possible. 
        
        
    \item (4) \emph{After rejection:} To compare the behavior of the website after initial rejection (opt-out) with that after the consent is explicitly revoked, we clean the browser (includes clearing history, cookies, cache, passwords, form data, and site settings to ensure consistency in each measurement) and record the options to reject consent on the website. If a consent banner is present, we reject all cookies. 
\end{description}
Amid all these steps, we also take screenshots in an automated manner at different stages by scrolling to the bottom of the pages or navigating to the settings in order to get the important information like keywords leading to revocation. 
Using these inputs and screenshots, we categorize the websites based on the revocation options, where such options are provided, and the number of steps required to revoke consent. 
We then map these categories to the legal requirements for GDPR and consent revocation (from \S\ref{sec:legal_bg}) to identify and measure potential violations of the law. 
%
To reduce manual bias in our analysis, labeling was conducted by one author (twice, three months apart, in March and June of 2024.), with category examples co-developed with a legal scholar. Several examples were jointly reviewed by both a legal expert and a computer scientist.
\label{sec:meth_cookies}
\subsubsection{Effect of revocation on AA cookies (RQ2)}
In the background, we collect cookies stored in the browser at each of the four stages. 
Since only some cookies require consent (see \S\ref{sec:legal_bg}), we followed  Bouhoula et al.~\cite{Bouh-etal-24-USENIX} and consider that only cookies classified as \emph{Advertising} or \emph{Analytics} (AA) by  CookieBlock~\cite{automating_gdpr_violation_detection} require user consent. 
%
If we detect AA cookies upon initial landing, after rejection, or after revocation, we map it to legal requirements and identify a potential violation of the law. Similarly, we analyse whether the number of AA cookies increases after revocation w.r.t. other stages of our analysis.  

%
In this analysis, we do not include the websites without a banner because we need to compare cookies after explicit acceptance and after revocation. Websites without a banner do not allow a user to accept consent, and therefore, modifying consent on such websites would not mean ``revocation'' in legal terms.

\subsection{Datasets for RQ3 and RQ4}
\label{sec:meth-rq34}
In websites that allowed revocation, different methods were used to store the modified consent and communicate it over the network. 
To detect the presence of CMPs that implement the IAB TCF~\cite{TCF}, and therefore use a standardised format for storing consent (Transparency and Consent String or \tcs), prior works~\cite{do_cookie_banners_respect, cschecker} used the {\tcfapi} function provided by TCF.
We, therefore, performed an automated preliminary study to test this detection method. We also identify the detection methods for the use of OneTrust, which is the most popular CMP on Tranco top-20k websites~\cite{hills-2021-pets}. 


\subsubsection{Preliminary study} We performed an automated crawl querying for {\tcfapi}, a function that must be provided to use TCF's functionalities, on the top-200 domains~\cite{trancolist}.
Out of 158 websites that were reachable, we found \tcfapi\ on only 
32 (20\%) of websites
\footnote{Similar to prior work~\cite{do_cookie_banners_respect}, we found that 
\texttt{\_\_tcfLocator} is present on only 1 website where \tcfapi\ is not present. 
 We therefore do not test for this method here.}. 
%
We therefore further explored 
other ways to 
detect CMPs by analysing the implementations of  
the most popular CMP~\cite{hills-2021-pets}, OneTrust. By  examining 
websites with OneTrust 
and its documentation~\cite{onetrust-dev}, we found that this CMP (1) stores user consent in a very specific format in a cookie named {\optanon}, and (2) maintains a JavaScript variable, \texttt{OneTrustActiveGroups} (\onetrustg), which can be queried to get the current consent string. 
However, the format of the consent string 
accessible through these methods
does not respect the official IAB Europe TCF {\tcs} format. 



\subsubsection{Website selection}
We use the methods developed in 
the preliminary study to detect websites that use TCF and OneTrust from top-200 websites, 
resulting in 
dataset \textbf{DS2} with only 56 websites with CMPs. We randomly chose an additional 1000 websites between the rank 200 and 5000 and detected 225 websites with either TCF or OneTrust 
resulting in a set of 281 websites (\textbf{DS2} and \textbf{DS3}). 

\begin{figure*}[ht]
  \centering
  \includegraphics[width=0.95\linewidth]{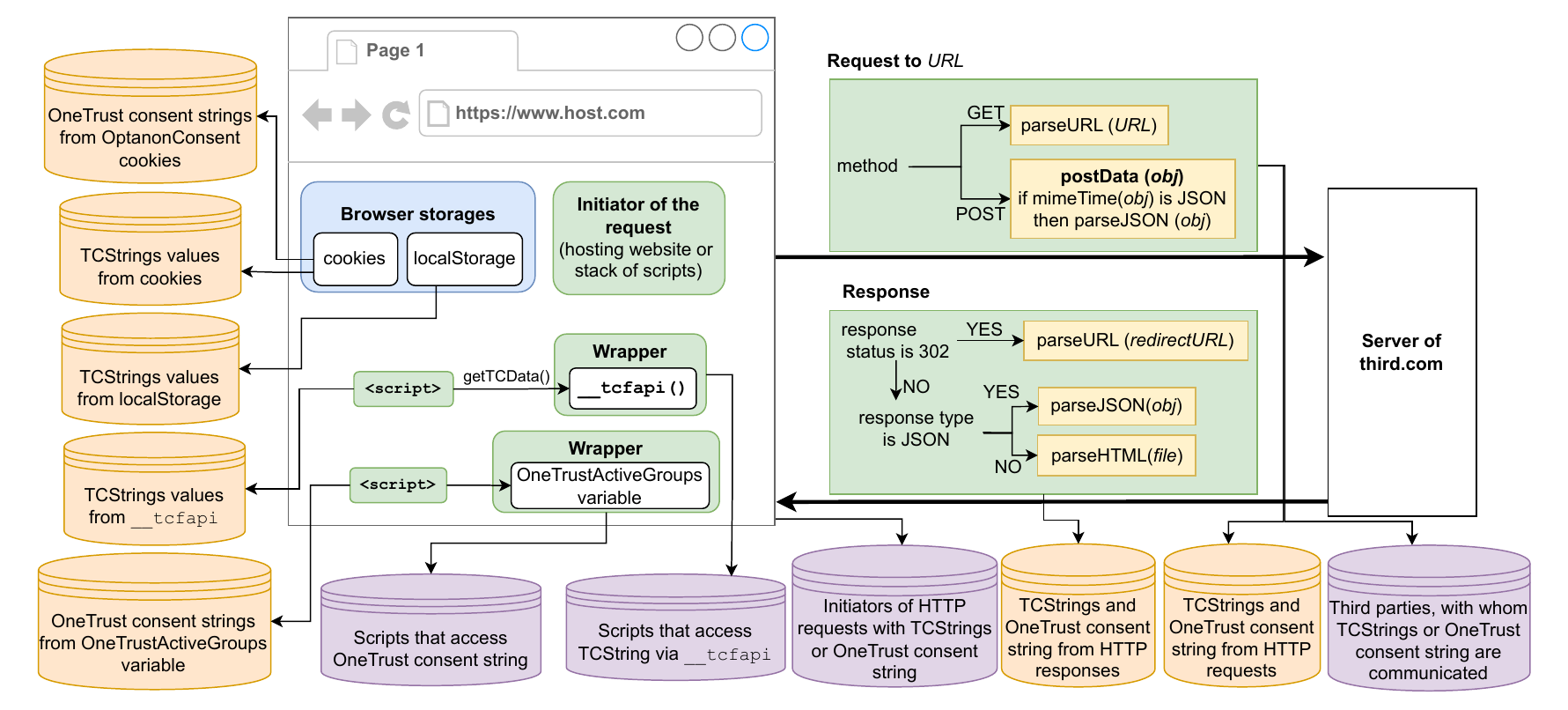}
  \caption{Data collection pipeline: to address \textbf{RQ3} (validity and consistency of consent), the collected sets of consent strings are highlighted in orange, while for \textbf{RQ4} (access and communication of consent to 3rd-parties), the sets of scripts parties involved in the communication of consent are highlighted in violet. The objects in green represent our methods for collecting the data, and the boxes in yellow are further described in Table~\ref{tab:parsing}.}
  \label{fig:data_analysis}  
\end{figure*}

\subsubsection{Data collection}
\label{sec:meth-data-collection}
Figure~\ref{fig:data_analysis} shows the data collection pipeline for \textbf{RQ3} and \textbf{RQ4}. The data is collected during 
all four stages of the experiment 
(highlighted as ``Save background data'' in Figure~\ref{fig:data_collection}). 
We collect the consent strings (both {\tcs} and consent variable specific to OneTrust) for further analysis of validity and consistency of consent storage (\textbf{RQ3}), and complement it with the information about scripts that access consent strings via APIs, scripts that initiate the requests and third-parties with whom consent is shared for the analysis of communication of consent (\textbf{RQ4}).

\subsubsection{Collecting stored consent strings and third-parties that access them}
\label{sec:meth_inconsistency}
The current TCF v2.2, does not define the specific location where the user consent should be stored. Therefore, we collect all cookies and localStorage objects' values that match the {\tcs} format. Additionally, we collect the {\optanon} cookie that OneTrust uses to store a user's consent. We also inject scripts in the webpage to query {\tcfapi} and {\onetrustg} to analyse consent strings that third-parties would receive if they query these APIs and examine network logs to detect and store {\tcss} shared in the requests and responses. All the collected consent strings are visualised in Figure~\ref{fig:data_analysis} in orange color. 

Third-parties can access consent strings via APIs or receive them via network requests. 
We capture  
(1) third-party scripts that fetch  consent string via {\tcfapi} or {\onetrustg}; 
(2) HTTP requests that contain consent strings in {\tcs} format or {\onetrustg}. However, some requests might not have been recorded since they do not have a well-defined format.
We store all initiators and receivers of consent strings (see violet stores in Figure~\ref{fig:data_analysis}).

\subsubsection{Detecting and extracting {\tcss} and specific  categories from OneTrust  in network logs} 
\label{sec:meth-logs}

\NBcom{Reviewer A: "Sec. 4.2.5: is this a violation when the third-party rewrite the TCString? If you uncover more than prior work, you need to say by how much (make a proper comparison). Here, the writing does not say anything about prior work numbers." Meta-review says "Further emphasize your results and how they differ from prior work"}

To detect the {\tcs} in network logs, we look at different parts of the request/response pair. 
%
Previous works have extracted {\tcss} only from specific URL parameters \NBtext{of only  outgoing HTTP requests}, such as  \texttt{gdpr\_consent}~\cite{do_cookie_banners_respect} following TCF v1.1. 
A follow-up work~\cite{cschecker} checked for outgoing {\tcs} 
in all URL parameters of HTTP requests and also in the HTTP cookie headers since TCF v2.2 does not specify URL parameters to be used. 
\NBtext{Differently from previous works, our approach is the first to compare TCStrings sent in outgoing HTTP requests sent to third-parties to the TCStrings returned in HTTP responses, identifying cases when third-parties \emph{modify the original TCStrings}.}

When parsing URLs, we extract all query parameters, for example for a URL \texttt{https://www.site.com /zzz.jpg?ISBN=XXX\&\-UID=ABC123}, we first extract the values \texttt{XXX} and \texttt{ABC123} and then check whether they correspond to the {\tcs} format, according to TCF specification~\cite{TCF_consentstring}. 
The orange boxes in Figure~\ref{fig:data_analysis} represent the functions we used to extract {\tcss} from different parts of the network logs that are further explained in Table~\ref{tab:parsing}.

Since {\tcss} do not have any protection mechanisms, any third-party that receives a {\tcs}, can potentially modify it and return a different {\tcs} back to the browser. To detect inconsistencies in {\tcs} between requests and responses, we analyse \texttt{postData} in 
POST requests,  
and the response data (HTML and JSON formats) from the servers. 
In each case, we record the request initiator to identify scripts that sent the wrong consent strings.  
To analyse HTTP GET requests, we search for {\tcs} in the URL parameters, while for POST methods, 
we extract the {\tcs} from JSON object if the data is sent in this format. To analyse  HTTP responses and detect when {\tcs} is modified by the server, we analyse redirected URLs: if response type is JSON, we parse it; otherwise, we parse the returned HTML file to extract {\tcs} from the script, image and iFrame elements. 
%

Our approach uncovers more instances of {\tcss} as compared to previous works~\cite{do_cookie_banners_respect,cschecker} --- we detect 4,445 {\tcss} in URL parameters, 3,423 {\tcss} in the \texttt{postData}, 52 {\tcss} in JSON objects and 170 {\tcss} in HTML files. 
%
%
%
We also record requests where the cookie containing {\tcss} is sent to third-parties.

\begin{table}[tb]
\begin{tabular}{|p{2cm}|p{5.6cm}|}
\hline
\textbf{Func. in Fig. \ref{fig:data_analysis}} & \textbf{Extracting \tcs} \\ \hline
parseURL (\textit{URL})
    & Extract all query parameters from \textit{URL} and keep all values that match a {\tcs} format or contain  OneTrust cookie categories.
\\ \hline
parseJSON (\textit{obj}) 
    & Extract all \{\texttt{key:value}\} pairs present in \textit{obj} and keep all values that match a {\tcs} format or contain OneTrust cookie categories.
\\ \hline
parseHTML (\textit{file}) 
    & Extract URLs for \texttt{img}, \texttt{iframe} and \texttt{script} in \textit{file} and parse them to extract {\tcss} or OneTrust cookie categories.
\\ \hline
\end{tabular}
\caption{Extracting {\tcss} and OneTrust consent string}
\label{tab:parsing}
\end{table}

\subsubsection{OneTrust-specific encoding of consent} 
Differently from other CMPs, OneTrust CMP does not use the TCString but uses a specific format to store and communicate consent: the consent can be stored as key-value pairs in two locations, either in the \optanon\ cookie or in the {\onetrustg} variable. 
The cookie values contain encoded categories of cookies (e.g., ``analytics'' or ``advertisement''), that follow the format such as ``X:1;Y:0;Z:1'', where X, Y and Z indicate cookie categories and 0 or 1 indicate rejection or acceptance to use the specific category in the consent.
The values of the {\onetrustg} list all the allowed cookie categories (called ``active groups''  in OneTrust documentation); however, the encoded format of such categories is not specified in OneTrust documentation. 
%
Our analysis of inconsistency between the \optanon\ cookie value and {\onetrustg} variable (section \ref{sec:consistency}) revealed that when two values are not consistent, \optanon\ cookie value contained correct record of user consent.
We therefore opted for extracting the allowed categories from the \optanon\ cookies and search all such encoding (only X and Z in the example above) in the URL query parameters to record the consent strings being sent to third-parties.

However, the variable in the query parameter having the consent string is not consistent. This leads to false positives where the active groups are represented using numbers like `1', `2', and `3' or ``1:1,2:0,3:0,4:1". These representations are hard to distinguish from query parameters used for purposes other than sharing consent. We manually analyzed such websites, and hence, could have missed some websites due to human error. 



\subsubsection{Wrappers for {\tcfapi} and {\onetrustg}} To identify {\tcfapi} access, we override the function to record all the scripts that access it. We also record what command was requested, e.g., ``ping'', ``getTCData'' or ``getEventHandler''. 
A similar approach was adopted by Matte et al.~\cite{do_cookie_banners_respect}. However, they observed calls made to the APIs in an older version of TCF, and their analysis did not include OneTrust. We override getter and setter methods of the {\onetrustg} variable to record which domain is writing/updating it (mostly by a designated CMP script) and reading/accessing it.

\subsubsection{Communicating consent modification}
We check the network logs to investigate if third-parties that received the {\tcs}and One Trust consent string (containing allowed cookie categories) after acceptance had also received the updated consent after revocation. We do this by observing requests to third-party URLs and if the request contained the {\tcs} or One Trust consent string either in the URL or in the POST data as described in Figure~\ref{fig:data_analysis}. 

\subsubsection{Identifying responsible parties}
We record the initiators of all the network requests and use this information to identify which third-party script changed the consent. We match the script's domain of provenance with the CMP name decoded from the {\tcss}. If the name and domain do not match, in case of inconsistencies, we manually check the domains for cases where the CMP names are either shortened or have words like `privacy center' or `CDN'. If we do not identify any such indicator that the third-party could be the CMP, we classify the responsible party just as a third-party. 

\subsection{Limitations}

\subsubsection{Accuracy of AA cookie classification} Even though CookieBlock~\cite{automating_gdpr_violation_detection} has better accuracy than other tools for classifying cookie purposes, it might not be entirely accurate: a website may declare some of the cookies, classified as AA, as \textit{necessary} in their cookie policies. 
Such cookies, classified by CookieBlock, will be, nevertheless, labeled as AA and considered to require consent.
%
%

\subsubsection{Not capturing network logs while searching for revocation option on secondary pages} When reaching the revocation settings takes multiple steps, for a few seconds, log-entries with the old consent (after acceptance) are also saved along with the network logs meant for ``after revocation'' stage. 
We remove these log-entries by searching for inconsistencies after the consent is revoked. However, the set of log-entries misclassified as ``after revocation'' are not checked for inconsistencies, and we may have missed a few cases similar to the ones mentioned in Table~\ref{tab:anomaly2} in the Appendix.

\subsubsection{Unable to capture access to consent when the third-parties use event listeners} Callback functions returned as an event listener by {\tcfapi} do not have a standard format or name to track the access requests made to the API. 
\NBcom{I propose to comment out the sentence below to gain space}
\NBcom{Reviewer A: "Sec. 4.3.3: But don't the event listeners have to initiate a network request to the third party anyway, which would then be captured? What is the point of event listeners without HTTP revocation notification after?" Our current text does not answer the question of seeing initiated HTTP request, we need to clarify the answer here. I added some text below.}
\NBtext{In theory, event-listeners should notify third-parties upon consent modification by initiating a network request. However, we observed that even when the third-parties registered the event listeners, there were no explicit network requests transmitting the modified consent string.}
Hence, we do not track API accesses when event listeners are used. 

\subsubsection{Determining description of purposes in OneTrust} We do not know how the purposes displayed by OneTrust or any CMP used by the website are mapped to the user's actual consent. Since there is no particular format used to store the consent string, it can vary between different CMPs and within websites using the same CMP. 

\subsubsection{Delay in registering user's consent} 
While collecting network logs to assess the communication of consent revocation to third parties, we observed (see Section \ref{sec:anomalies}) that some websites may take time to register that a user revoked their initial consent.
To account for this, we exclude cases with delays in updating consent strings from our list of violations.
%
%
However, we include cases where some third parties correctly include the updated consent string in their network requests, while others still send the old consent (positive consent string). 
As the third-party notification happens server-side, this may result in some false-positives on our end.



\subsection{Data Availability and Disclosure}
\label{sec:data-disclosure}
As described in \S\ref{sec:meth-data_collection} and Figure~\ref{fig:data_collection}, we collect background data, screenshots, and website labels. The complete dataset, including the network logs and the crawler, is provided in the supplementary material~\cite{sup-material}. Some examples are shown in the Appendix. 



As of the date of submission of this paper, 28 February 2025, we have notified all the companies who own the domains that we explicitly mention in the main text of our paper. 
We will add information on the responses we receive, in the final draft. 

\section{Revocation Interface and its Compliance}
\label{sec:interface}



To address \textbf{RQ1}, we examined user interfaces for revoking consent and checked if they (potentially) violated EU legal requirements.
%
%
We first categorize the $158$ reachable websites (datasets DS1 and DS2 from Table~\ref{tab:datasets}), 
based on their consent banners:
108 (67\%) websites display a \emph{consent banner} to the users on their first visit,
 8 (5\%)  websites displayed a banner with \emph{no option}, not allowing to accept or reject consent, and 
 45 (28\%) websites did not display any banner.\footnote{This can be explained by the fact that these websites were hosted outside of the EU and did not yet implement  compliance solutions with EU laws.} 
We manually classify the 158 websites based on the interface to revoke consent: 
120 (74.5\%) websites provide users option to revoke consent \emph{within the same interface} where the consent request took place, or navigates to related pages for revoking consent; 
32 (19.8\%) websites provided options \emph{via different interface}, where such options are present outside the interface or medium, where the consent request took place, and 
9 (5.6\%) websites offer \emph{no revocation} option.  

\newcommand{\tgreen}[1]{\cellcolor[HTML]{D9EAD3}{#1}}
\newcommand{\tyellow}[1]{\cellcolor[HTML]{FFF2CC}{#1}}
\newcommand{\tdred}[1]{\cellcolor[HTML]{EA9999}{#1}}
\newcommand{\tred}[1]{\cellcolor[HTML]{F4C2C2}{#1}}

\begin{table*}[ht]
\centering
\begin{tabular}{|l|r|r|r|r|r|r|r|r|r|}
\hline
& & \multicolumn{4}{c|}{\textbf{Within the Same Interface}} & \multicolumn{3}{c|}{\textbf{Via Different Interface}} & \multicolumn{1}{c|}{\textbf{No Revocation}}                                              
\\ \cline{3-10} 
\multicolumn{1}{|c|}{\multirow{-2}{*}{\textbf{Banner}}} & \multicolumn{1}{c|}{\multirow{-2}{*}{\textbf{\#}}} &
\multicolumn{1}{c|}{\tgreen{{\begin{tabular}[c]{@{}c@{}}Icon\end{tabular}}}} & 
\multicolumn{1}{c|}{\tgreen{{\begin{tabular}[c]{@{}c@{}}Footer\\Options\end{tabular}}}} & 
\multicolumn{1}{c|}{\tgreen{{\begin{tabular}[c]{@{}c@{}}Banner\\on Policy\end{tabular}}}} & 
\multicolumn{1}{c|}{\tyellow{{\begin{tabular}[c]{@{}c@{}}Options\\via Policy\end{tabular}}}} & 
\multicolumn{1}{c|}{\tred{{\begin{tabular}[c]{@{}c@{}}Settings\\or links\end{tabular}}}} & 
\multicolumn{1}{c|}{\tred{{\begin{tabular}[c]{@{}c@{}}After\\Login\end{tabular}}}} & 
\multicolumn{1}{c|}{\tred{{\begin{tabular}[c]{@{}c@{}}Contact/\\Email\end{tabular}}}} &
\multicolumn{1}{c|}{\tdred{{\begin{tabular}[c]{@{}c@{}}No option\\provided\end{tabular}}}} 
\\ \hline
Consent banner & 105 & \tgreen{8} & \tgreen{61} & \tgreen{4} & \tyellow{23} & \tred{8} & \tred{1} 
& \tred{0} & \tdred{0}                       
\\ 
No option banner & 8 & \tgreen{0} & \tgreen{0} & \tgreen{0} & \tyellow{5} & \tred{3}& \tred{0}
& \tred{0} & \tdred{0}                                       
\\ 
Without banner & 45& \tgreen{0}& \tgreen{7}& \tgreen{2}& \tyellow{7}& \tred{17}& \tred{0}& \tred{3}
& \tdred{4 {[}+5{]}}
\\ \hline
\multicolumn{1}{|c|}{\textbf{Total}} & 158& \tgreen{8}& \tgreen{68}& \tgreen{6}& \tyellow{35}& \tred{28}& \tred{1}& \tred{3}
& \tdred{9}
 \\ \hline
\end{tabular}
  \caption{Prevalence and type of consent revocation options on 158 websites (rank 1-200) from \textbf{DS1}. We use different colors to represent the level of compliance of the detected practices. 
  Green color represents zero or one steps, yellow represents two or more steps, and pink and red represent violations of varying severity. 
  }   
  \label{tab:data_analysis1}  
\end{table*}

Table~\ref{tab:data_analysis1}  summarizes the results for these three categories. 
To further evaluate legal compliance, for websites that provide revocation \emph{within the same interface}, we  count the number of steps required to revoke consent based on Figure~\ref{fig:data_collection}. 
 %
Figure~\ref{fig:within_rev} (in the Appendix) shows example screenshots for each of these categories.





\subsection{Results}
\subsubsection{Compliant revocation interface (zero or one steps)}

\label{sec:interface_1}

Out of the 158 websites,
only 8 (5.6\%)  offer 
a persistent icon or button floating on the page, thus requiring \emph{zero steps} to reach the revocation option. 
%
$68$ 
(41.6\%)
websites offered a link option in the footer of the page, 
requiring \emph{one step} to reach the revocation option. 
6 
(6.8\%)
websites showed a consent banner or icon when accessing the privacy policy page from the footer, therefore also requiring \emph{one step} to revoke consent. 
All these implementations found on 82 (51\%) websites (labeled as \emph{Icon, Footer Options} and \emph{Banner on Policy} in Table~\ref{tab:data_analysis1}) \emph{comply} with revocation  GDPR principles and consent requirements (\textbf{\hyperref[lr1]{LR1}-\hyperref[lr3]{3}}) since they are presented within the same interface and require zero or one step to revoke consent. 
According to Habib et al.~\cite{Habib-etal-2022}, users who face persistent icons are more likely to recognise a correct method to revoke consent with respect to users who saw a link  in the website's footer. Therefore, even though all such  implementations are compliant, only 7.4\% websites with an \emph{Icon}  provide a more \emph{usable} revocation design. 

\subsubsection{Two or more steps  to revoke vs zero steps to accept}
 35  (22.1\%) websites out of 158 allow users to access the revocation option \textit{within the same interface}, but with additional obstruction, requiring them \emph{2 or more steps} to revoke consent. Such websites (labeled with \emph{Options via Policy}) hid the option to revoke consent inside the ``Cookie Policy'' or ``Privacy Policy'' page.
%
According to the majority of EU regulators, 
this additional effort does not allow to exercise the right of revocation in a easy and effective way even if such option is located withing the interface. Arguably,  
the websites requiring additional effort  neither comply with \hyperref[lr3]{\textbf{LR3}} (Easy revocation through the same effort and number of steps)
nor with the principles \hyperref[prin1]{\textbf{P1}} 
(Fairness), since websites require unjustifiable and unexpected effort, nor \hyperref[prin2]{\textbf{P2}} 
(Data Protection by Design), since the adopted measures are not efficient as to facilitate the revocation right but obstruct it instead, and \hyperref[prin3]{\textbf{P3}} (Accountability), since such websites are not able to demonstrate compliance with the 
requirement \hyperref[lr3]{\textbf{LR3}}.

\subsubsection{Revocation options \textit{Via Different Interface}} 
Overall, 32 (19.87\%) websites out of 158 offered the option to revoke consent via an interface that is substantially different from the interface for users to accept consent.
Figure~\ref{fig:browser_rev} in the Appendix shows examples of websites for each of the options given. 

Twenty-five websites, including \texttt{medium.com, discord.com} and \texttt{wikipedia.org}, suggest revocation through \emph{browser settings} by clearing cookies  
or by offering opt-out links for  third-party tracking or advertisement domains 
(labeled as \emph{Settings or links}).  
Three websites -- \texttt{github.io, archive.org} and \texttt{who.int} -- suggested users to contact or email them to revoke consent or delete the data (labeled \emph{Contact/email}).

\textit{Settings or links} as well as \textit{Contact/email} revocation options infringe the legal requirement \hyperref[lr2]{\textbf{LR2}}  
 (Easy revocation through the same interface),
since revoking consent is not made available through the \emph{same} means or interface; and principles \hyperref[prin1]{\textbf{P1}} 
and \hyperref[prin2]{\textbf{P2}}, 
since switching to these totally different interfaces requires unnecessary and disruptive effort that is obstructive and unexpected. This leaves users in an asymmetrical relationship between giving and revoking consent, which does not permit users to exercise their revocation right. Consequently, these websites cannot be accountable for demonstrating compliance with
this right (\hyperref[prin3]{\textbf{P3}}).


On one website(\texttt{tumblr.com}) privacy settings lead the user to the login page and instruct them to revoke consent \emph{after logging-in}, which may require account creation (labeled \emph{After login}). 
%
%
While this option 
occurs through the same website, it directs the user to a different interface. As before, this option infringes the requirement \hyperref[lr2]{\textbf{LR2}}  
since the option redirects users to an unrelated interface. It also infringes principles  \hyperref[prin1]{\textbf{P1}} 
and \hyperref[prin2]{\textbf{P2}} 
as it forces users to take unexpected additional steps and effort -- 
to subscribe, and thus to give more personal data -- to revoke consent. This option impedes these websites to show accountable revocation (\hyperref[prin3]{\textbf{P3}}). 
\subsubsection{No revocation option provided}
Nine (5.6\%) websites out of 158 did not provide any means to revoke consent, though they also did not display any consent banner, indicating that these websites probably did not integrate online tracking. 
We  further analysed these websites and found that 4 (2.48\%) of them stored AA cookies making them non-complaint. 
These websites include \texttt{ntp.org} and \texttt{un.org}.
All these 4 websites that use AA cookies 
 infringe the legal requirement \hyperref[lr1]{\textbf{LR1}}  
  (Right to revoke consent)
 and the following principles: \hyperref[prin1]{\textbf{P1}} 
 since websites process data unjustifiably for AA purposes without a legal basis, or the knowledge or consent of users; \hyperref[prin2]{\textbf{P2}} 
 as it does not create measures to enable the exercise of the revocation right. Consequently, these websites cannot be accountable for demonstrating compliance with this right (\hyperref[prin3]{\textbf{P3}}). 
 




\begin{figure}
  \centering
  \includegraphics[width=\linewidth]{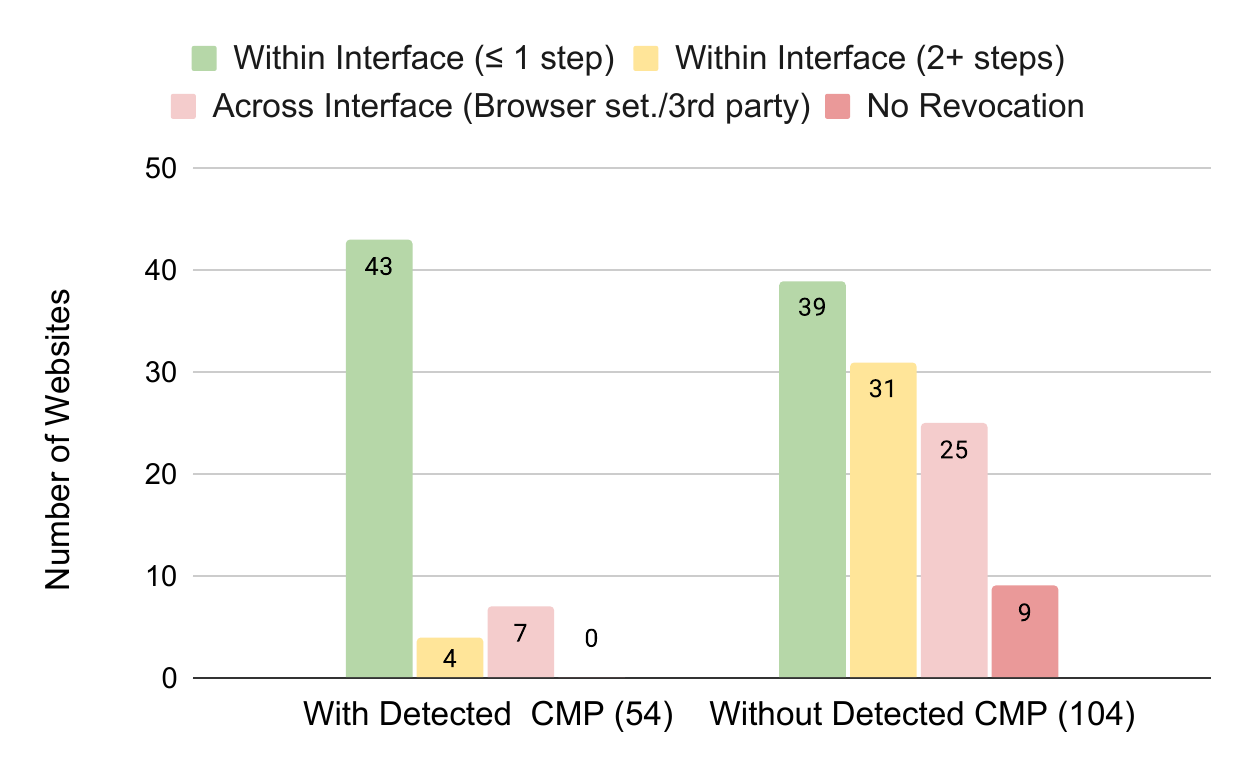}
  \caption{Revocation methods on top 158 websites}
  \label{fig:comparison}  
\end{figure}

\subsection{Impact of CMPs on compliance}
Figure~\ref{fig:comparison} shows the prevalence of different revocation options on the websites with detected CMPs \textit{vis a vis} websites where we did not detect any CMP. 43 (77\%) out of 56 websites  with detected CMPs provide a compliant implementation (zero or one step) to revoke consent, while only  32 (30.5\%) websites out of 105 websites without detected CMP have compliant implementation, showing overall a higher compliance rate for websites with CMPs. 

Regarding non-compliant implementation, 
37\% (39 out of 105) 
websites without CMP 
require two steps or more to revoke consent, while only
10.7\% (6 out of 56)
websites with CMP needed 2 or more steps for revocation. 
None of the websites with CMP denied a revocation option to users, though 9 out of 105 websites without CMP offered no revocation option.
%
In conclusion, websites with detected CMPs tend to be more compliant with revocation requirements.

\section{Effect of Revocation on AA Cookies}
\label{sec:cookies}
\begin{figure*}[ht]
  \centering   
  \begin{subfigure}[b]{0.31\textwidth}
    \centering
    \includegraphics[width=\textwidth]{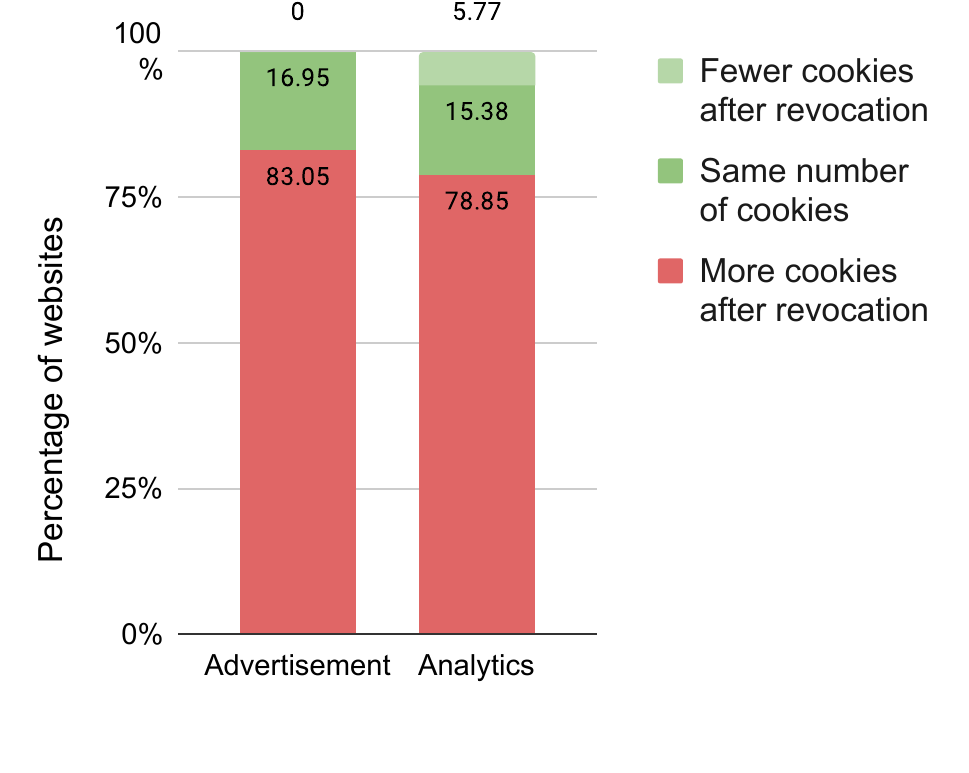}
    \caption{Comparison of AA cookies upon\newline
    \emph{initial landing} vs \emph{after revocation}}
    \label{fig:cookie_variation2}
    \end{subfigure}
  \hfill
    \begin{subfigure}[b]{0.31\textwidth}
      \centering
      \includegraphics[width=\textwidth]{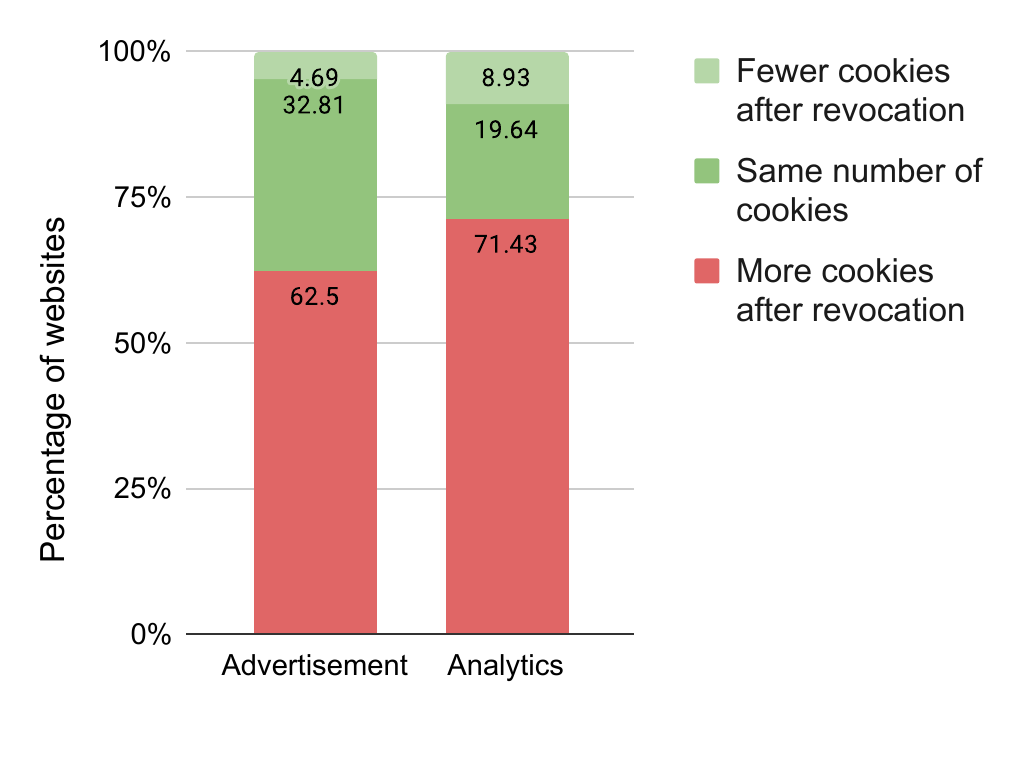}
      \caption{Comparison of AA cookies\newline 
      \emph{after rejection} vs \emph{after revocation}}
      \label{fig:cookie_variation3}
    \end{subfigure}
    \hfill
    \begin{subfigure}[b]{0.31\textwidth}
    \includegraphics[width=\textwidth]{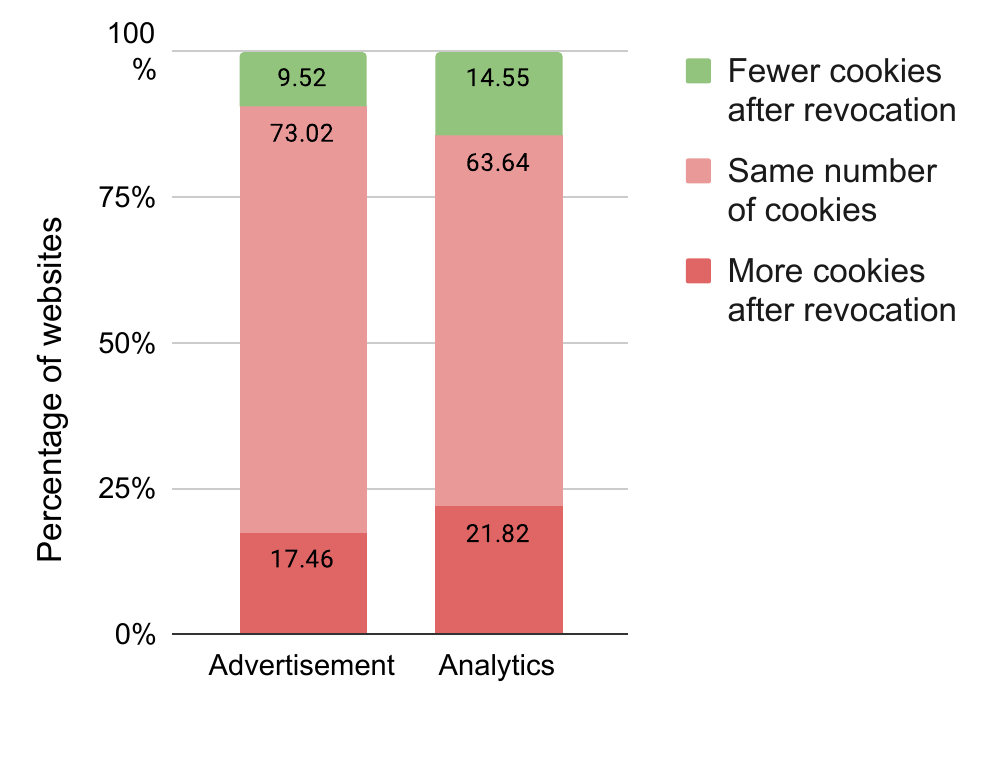}%
    \caption{Comparison of AA cookies \newline
    \emph{after acceptance} vs \emph{after revocation}}
    \label{fig:cookie_variation1}
    \end{subfigure}
    \hfill
\caption{Change in the number of advertising and analytics (AA) cookies across different options} 
\label{fig:cookie_variation}
\end{figure*}
  
In this section, we address \textbf{RQ2} and measure whether cookies requiring consent (Advertising and Analytics, or AA cookies) are deleted upon user's consent revocation. 
As described in \S~\ref{sec:meth_cookies}, we recorded the cookies of websites in four stages -- 
upon initial landing, after accepting consent, after rejecting optional cookies, and after revocation. 
%
%
%
%
Out of the 120 websites where revocation was possible within the same interface, we find AA cookies after revocation on 69 (57.5\%) websites. 

Figure~\ref{fig:cookie_variation} compares the change in the number of AA cookies  on websites after revocation w.r.t. other stages.
%
%
%
Surprisingly, the number of AA cookies increases on most  websites after revocation w.r.t. initial landing and w.r.t. after rejection (see red bar in Figures~\ref{fig:cookie_variation2} and \ref{fig:cookie_variation3}). 
%
%
Additionally, on the majority of analysed websites, the number of AA cookies after acceptance remains the same after revocation (see pink bars in Figure~\ref{fig:cookie_variation1}). 
Websites that \emph{add or keep} AA cookies after consent revocation violate \hyperref[lr4]{\textbf{LR4}}  (Revoking requires stopping data processing and deletion of consent-based data),
 \hyperref[prin1]{\textbf{P1}} 
 and \hyperref[prin2]{\textbf{P2}}. 
Such AA cookies are processed unexpectedly and contrary to user's decisions, without a legal basis and are thus illegal ~\cite[Art. 6(1)(a)]{gdpr}.

\section{Validity and Consistency of Consent}
\label{sec:consistency}

In this section, we analyse how consent information is stored and shared behind a website interface, whether it is consistently stored across different storage and APIs and whether it is legally valid, thus answering \textbf{RQ3}.
We analysed 281 websites with detected CMPs (DS2 and DS3 from Table~\ref{tab:datasets}) to check the validity and consistency of the consent string that these CMPs implement. 
Together with a legal expert co-author of the paper, we analyzed 11 purposes predefined within IAB Europe TCF v2.2~\cite{TCFv22} (see Table~\ref{tab:purposes} in the Appendix). These purposes are largely consistent with those of 
the  TCF v2.0~\cite{Matt-etal-20-APF}. 
We determined that purposes 2-9 require consent and demand user action to be selected, thus such purposes cannot be enabled by default in {\tcs}. 
Purposes 10-11 are exempted from this requirement and may be enabled by default. 
Purpose 1 is storage-based and 
enabled by default. 

 \noindent 
\textit{Positive consent:} if the given {\tcs} contains 
at least one of the purposes  from 2-9, and at least one vendor in its vendor list, we consider it to contain a positive consent. Such consent is correct only within the Acceptance phase, where the user has actively selected  purposes 2-9  requiring user action.

\noindent 
\textit{Negative consent:} if
only purposes 1, 10 or 11 are enabled in the {\tcs}, 
we conclude that it contains negative consent because none of these purposes require any user action as per our legal analysis. 
Consequently, if such {\tcs} is present upon \emph{initial landing}, or \emph{after rejection} or \emph{after revocation} stage, we consider consent to be registered correctly.
%

%
%



For the consent strings extracted from {\onetrustg} variable, or from {\optanon} cookie, 
we follow the OneTrust specification~\cite{onetrust-dev} and extract purpose numbers from its ``groups'' parameters. Since we do not have a specification for the meaning of these purposes, we cannot analyse which ones require consent. We  therefore assume that a consent string contains a \emph{negative consent} if it matches the value observed upon the \emph{initial landing} stage, and a \emph{positive consent} if the value contains more purposes than at \emph{initial landing}.  




\begin{table*}
\centering
\begin{tabular}{|l|r|r|r|r|r|r|} 
\hline
\multicolumn{1}{|c|}{\multirow{2}{*}{\textbf{Storage method or API}}} & 
\multicolumn{3}{c|}{\textbf{Number of websites}} & 
\multicolumn{3}{c|}{\textbf{Positive consent}}\\ \cline{2-7}

& 
\multicolumn{1}{c|}{\begin{tabular}[c]{@{}c@{}}Total using\\the method\end{tabular}} & 
\multicolumn{1}{c|}{\begin{tabular}[c]{@{}c@{}}Rejection\\possible\end{tabular}} & 
\multicolumn{1}{c|}{\begin{tabular}[c]{@{}c@{}}Revocation\\possible\end{tabular}} & 
\multicolumn{1}{c|}{\begin{tabular}[c]{@{}c@{}}Initial\\landing\end{tabular}} & 
\multicolumn{1}{c|}{\begin{tabular}[c]{@{}c@{}}After\\rejection\end{tabular}} & 
\multicolumn{1}{c|}{\begin{tabular}[c]{@{}c@{}}\textbf{After}\\\textbf{revocation}\end{tabular}} 
\\ \hline
{\tcfapi} & 163 & 150 & 136 & 1 & 2 & \textbf{17} (15 not updated) 
\\ \hline
TCF Cookies & 130 & 104 & 101 & 0 & 0 & \textbf{12} (11 not updated) 
\\ \hline
TCF localStorage & 41 & 31 & 32 & 0 & 1 & \textbf{8} (7 not updated) 
\\ \hline
{\onetrustg} & 176 & 164 & 152 & 0 & 10 & \textbf{22} (16 not updated) 
\\ \hline
\optanon\ Cookies & 131 & 130 & 130 & 0 & 0 & \textbf{13} (10 not updated)  
\\ \hline
\end{tabular}
\caption{Websites where consent strings in browser storage and returned by APIs were positive in different phases.} 
\label{tab:empty-consent}
\end{table*}

\subsection{Results}
\subsubsection{Positive consent upon initial landing, after rejection and after revocation}
We first analyse if the {\tcss} returned by {\tcfapi} can be considered as a ``baseline'' in our comparison, i.e., whether such {\tcss} are valid. For this purpose, we  analysed whether the {\tcfapi} returns a positive consent at three stages: Initial landing, Rejection and Revocation, as shown in Table~\ref{tab:empty-consent}. 
%

%
\emph{After Rejection and Initial landing:}
Only two prior works~\cite{do_cookie_banners_respect,cschecker} analysed whether positive consent is stored or returned by APIs upon Initial landing and After Rejection. 
Compared to previous works, we found very few cases where positive consent is present after rejection in websites implementing the TCF (only 3 websites out of 150). In contrast, 
10 websites, out of 164, store positive consent \emph{after rejection}  with {\onetrustg}. 

\emph{After Revocation:}
We detected multiple websites where a \emph{positive consent is present after revocation}, which has not been studied in previous works. 
In the websites with the TCF,
17 (12.5\%)  out of 136
websites provide positive consent in the \tcfapi,
 12 (11.9\%) out of 101 websites
 store a positive consent in a cookie, and 
 8 (25\%) out of 32 
 websites store a positive consent in a localStorage. 
In the websites using OneTrust, we observe 
 22 (14.5\%) out of 152 websites with positive consent returned by {\onetrustg} , and 
 13 (10\%) out of 130 websites with positive consent in the \optanon\ cookie. 
Positive consent constitutes potential  violations of \hyperref[lr5]{\textbf{LR5}}  (Correct Consent Registration) 
which entails that the registered consent must be identical to the user’s choice in the user interface; and of the principle \hyperref[prin2]{\textbf{P2}} 
 since websites did not implement technical measures and safeguards to assure that revocation is done efficiently.






\subsubsection{Consent strings are often not updated after revocation} Since positive consent persisted post-revocation in many cases, we examined whether consent strings were properly updated (Table~\ref{tab:empty-consent}). Among websites using {\tcfapi}, 15 (9.3\%) did not update the consent string, while 11 (10.9\%) and 7 (21.9\%) failed to update the {\tcs} in cookies and localStorage, respectively. Notably, \texttt{sourceforge.com} lacked a banner, initially stored negative consent, but changed it to positive upon revocation. \texttt{ft.com} updated only the “consentScreen” parameter, keeping positive consent intact. For websites using \onetrustg, 16 (10.5\%) out of 152 failed to update consent post-revocation. Additionally, six websites modified only some purposes while still retaining positive consent. Interestingly, \texttt{cisco.com}, \texttt{opendns.com}, and \texttt{webex.com} added one purpose to the consent string after revocation.

\subsubsection{Inconsistency among consent strings between browser storage and APIs} 
Table~\ref{tab:storage-inconsist} shows the number of websites where the {\tcs} found in the cookies and localStorage is not consistent with the {\tcs} returned by the {\tcfapi}, and where the {\optanon} cookie does not match the consent value returned by the {\onetrustg}. 
Such mismatches result in either an incorrect storage of consent or incorrect functioning of the APIs, and result in wrong consent being returned to third-parties. 
%
%
%
%
5 websites implementing the TCF returned different {\tcss} from {\tcfapi} and the {\tcs} stored in the cookie after revocation. 
In 2 of such websites, (\texttt{freep.com} and \texttt{megaphone.com}) the {\tcs} in the cookie 
was not updated, and still contained a positive consent, while the {\tcs} returned by the {\tcfapi} was properly updated into a negative consent. 
In the website \texttt{aol.com}, while the {\tcfapi} returned a 
positive consent, the value in the cookie stores 
negative consent. 

%
In three websites (\texttt{reuters.com}, \texttt{manch\-est\-er\-even\-ing\-news\-.co\-.uk} and \texttt{portfolio.com}), 
the {\tcs} returned by {\tcfapi} showed negative consent, while the {\tcs} stored in the localStorage remained unchanged, i.e, it contained  positive consent. 
%
On websites implementing OneTrust, we observed 4 mismatches between {\onetrustg} and {\optanon} cookie value, where the cookies stored a negative consent and were shared on the network.  
This mismatch can lead to the incorrect consent being shared to third-parties. 

These mismatches between consent strings across browser storage and APIs can be due to the fact that websites do not consistently update the storage and APIs when users revoke consent. Consequently, websites do not correctly register consent, as they are obligated to, 
thus, infringing the consent requirement \hyperref[lr5]{\textbf{LR5}}. Moreover, \hyperref[prin2]{\textbf{P2}} 
is not also complied with since websites do not implement technical measures to assure that revocation is done efficiently.




\begin{table}
\centering
\begin{tabular}{|p{4.8cm}|p{2.7cm}|} 
\hline
\textbf{Browser storage vs. API}
& \textbf{\# websites}
\\ 
\hline
TCF cookie vs. {\tcfapi} & 5 (100) \\ \hline
TCF localStorage vs. {\tcfapi} & 3 (32) \\ \hline
OT cookie vs. {\onetrustg} & 4 (130) \\ \hline
%
\end{tabular}
\caption{Inconsistent consent strings across  browser storages and APIs after revocation. The numbers in brackets show the number websites that used both the storage and the API.}
\label{tab:storage-inconsist}
\end{table}

\subsubsection{Inconsistency between consent returned via {\tcfapi}  and consent shared on the network}
We 
compare consent strings from {\tcfapi}  with the consent strings found in outgoing network requests and incoming responses for two stages: Acceptance and  Revocation. We consider the cases where the {\tcs} returns the correct consent, i.e, a negative consent in case of revocation. 

We investigate  inconsistencies in the consent strings shared to and received from the third-parties, which, in turn, helps identify the responsible party for such inconsistencies. Previous works have only partially analysed such inconsistency~\cite[\S VIII.A]{do_cookie_banners_respect}, where positive consent was found to be sent within a specific \texttt{gdpr\_consent} URL parameter on websites where {\tcfapi} did not contain a positive consent string. We however observed that the consent can be shared via differently named URL parameter as well. Additionally, the consent string can be sent in the POST data as part of the request as well. We find different inconsistencies on 8 distinct websites out of the 136 websites where revocation is possible. 

These inconsistencies could arise due to: (1) delay in updating the consent after user revokes consent; (2)  some scripts not being updated about revocation in consent; (3) introducing/using a different \tcs. We do not consider delay in sending the updated consent as a violation since it is implementation specific. However, there are 8 websites where we observe inconsistencies causing possible violations. Out of these 8 websites, 4 of them (\texttt{forbes.com, time.com, n-tv.de} and \texttt{cadenaser.com}) had a different {\tcs} on the network while \texttt{deadline.com, kotaku.com, manch\-est\-er\-even\-ing\-news.co.uk} and \texttt{walesonline.co.uk}, we observed that the old positive consent string was shared on the network by some scripts even after revocation. Details regarding these inconsistencies can be found in Appendix~\ref{sec:anomalies}.


The detected mismatches between consent strings returned by {\tcfapi} and the {\tcs} shared on the network (Table \ref{tab:anomaly2}) lead to an incorrect registration of consent revocation by  websites and also by the implemented CMPs. 
Consequently, websites do not comply with \hyperref[lr5]{\textbf{LR5}} since they do not correctly register user revocation, as they are obligated to, in order to assure that the registered consent is identical to the user's choice in the user interface. As a result, a single {\tcs} does not serve as a proof of revocation, which should consist of a negative consent. 
Moreover, the  principle \hyperref[prin2]{\textbf{P2}} 
is also not complied with since websites did not implement technical measures to assure that revocation is done efficiently.

\section{Communicating Consent Revocation to Third-parties}
\label{sec:inform-consent-revocation}

Next, we address \textbf{RQ4} and investigate if all the third-parties that were informed of the consent acceptance (either by accessing APIs or via  HTTP requests) are also informed of consent revocation. 
%

%

\subsection{Results}
\subsubsection{Not all third-parties are informed of revocation by accessing consent via APIs} 
\label{sec:informing-consent}
To examine which third-parties access the APIs implemented by CMPs, 
we override the implementation of {\tcfapi}  and {\onetrustg}
(see \S\ref{sec:meth_inconsistency}).
%
%
%
%
Overall, we found 23 (9.6\%) websites out of 238 where at least one third-party accessed the API to fetch positive consent after acceptance, but did
 not access the API to fetch revoked consent. 
Among these, on 163 websites where {\tcfapi}  is implemented (see Table~\ref{tab:empty-consent}), 
on 14 (8.6\%) websites at least one third-party requests the {\tcfapi} with the command ``getTCData'' after acceptance but not after revocation\footnote{In the latest version of TCF, there is an option for third-parties to add an event-listener for consent update. However, since the implementation of this API is not standard, we could not intercept calls to it.}. 
%
%
%
%
Out of 176 websites that support {\onetrustg}, on 13 (7.4\%) websites at least one third-party accesses the {\onetrustg} variable after accepting, but not after revocation. 
On 4 of these 13 websites, scripts from the third-party domain \texttt{ads-static.conde.digital} read the value of the variable, indicating that this variable can be used to know the consent choice given for AA purposes, since this domain is present in EasyList~\cite{easylist}. 
%
%
%

CMPs expose APIs to third-parties to access consent information making them responsible for requesting access to consent and for being informed on updates. 
Therefore, a proper implementation of event listeners by the website or by CMPs is necessary to inform third-parties about consent revocation, since consent can be updated multiple times in a single user session. We discuss about the standardization of the event-listeners in \S\ref{sec:disc}.
\begin{table}[tbp]
\centering
\begin{tabular}{|p{4.5cm}|p{3cm}|}
\hline
\textbf{\% 3rd-parties not informed} & \textbf{Number of websites}\\ \hline
$< 25$\%                               & 1                                                                          \\ \hline
$\geq 25$ to $< 50$\%                            & 5                                                                          \\ \hline
$\geq 50$ to $< 75$\%                            & 15                                                                         \\ \hline
$\geq 75$ to $< 100$\%                            & 35                                                                         \\ \hline
$100$\%                                & 45                                                                         \\ \hline
\end{tabular}
  \caption{Percentage of third-parties 
  informed of acceptance but 
  not informed of the revocation in 101 websites. }
  \label{tab:tp}  
\end{table}



\subsubsection{Not all third-parties are informed of the consent revocation via HTTP requests}
%
On each website where consent revocation is possible, we detected third-parties that are informed of consent information via the {\tcss} and One Trust consent string in the URL or postData (see \S\ref{sec:meth_inconsistency}).  We then detect third-parties informed of consent after acceptance but not informed of revocation.

\begin{figure}[tb]
  \centering
  \includegraphics[width=\linewidth]{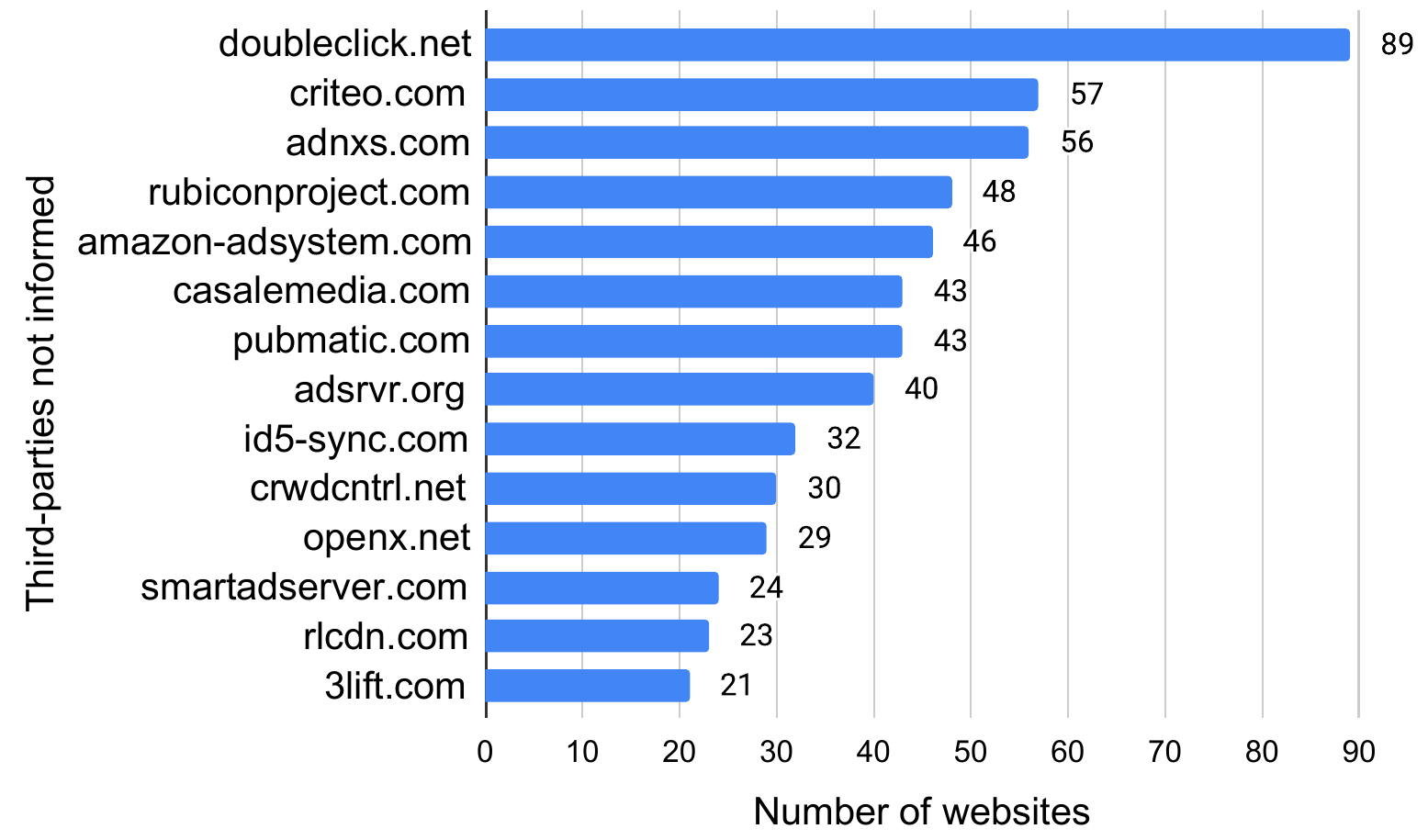}
  \caption{Top 15 third-parties not informed of the revoked consent via HTTP requests and the number of websites including these scripts but not informing them of revocation.}
  \label{fig:tp2-full}  
\end{figure}

Of the 136 websites that implement {\tcfapi} and support revocation, 101 (74.2\%) websites contain at least one third-party that was informed of consent after acceptance, but not after revocation via HTTP requests. Moreover, on 68 of these websites, third-parties that were not informed about revocation set cookies through the ``Set-Cookie'' HTTP header, and thus, did not stop processing user's data after revocation. 
Table~\ref{tab:tp} shows the percentage of third-parties not informed of  consent after revocation.
Surprisingly, 45 (44.5\%) websites did not inform \emph{any} of the detected third-parties, while 35 (34.6\%) of them did not inform more than 75\% of included third-parties. 
%
%
%
%
Figure~\ref{fig:tp2-full} shows 
most prevalent third-parties that were not informed  of the revoked consent via HTTP requests but were informed of consent acceptance. 
These websites include the big ad-tech industry actors \texttt{doubleclick.net}, \texttt{criteo.com}  and \texttt{adnxs.com}. 
Interestingly, 73.77\%  of all  third-party domains that do not receive a communication about consent revocation 
are present in EasyList~\cite{easylist} and therefore participate in advertising or tracking.

{
Of the 152 websites that provide revocation and implement OneTrust, we observed that allowed categories from the \optanon\ cookie (see \S\ref{sec:meth-logs}) are rarely shared with third-parties, i.e., 
only on 8 websites OneTrust allowed categories were sent to third-parties after acceptance, and 6 sent the allowed categories after revocation. 
Interestingly, all third-party requests with allowed categories were made to Google Analytics 
(\texttt{www.google-analytics.com}) and Google Tag Manager 
(\texttt{gtm.elementor.com}). According to the official OneTrust documentation, the categories are communicated to the Google Tag Manager  for it  to manage consent and third parties \cite{OT_GTM_integration}). We therefore conclude that OneTrust CMP may delegate the consent communication to the GTM framework and not send the allowed categories to third-parties by itself.
}

\subsubsection{Failure to communicate consent revocation results in unlawful data processing}
Our findings indicate that third-parties are not informed about consent revocation. 
This fact entails that these tracking- and advertising-based third-parties may continue to \emph{unlawfully}  process user's personal data even after users have revoked consent on the website's interface, since they were not updated of the user's decision by the website. 
Consequently, websites from both cases (communication of the {\tcs} via API and HTTP requests) are in potential violation with \hyperref[lr6]{\textbf{LR6}} (Communication of withdrawal to third-parties) 
that demands  websites to communicate consent revocation to the third-parties. The continuous  unexpected processing of user's data after revocation infringes the fairness principle \hyperref[prin1]{\textbf{P1}} and the data protection by design principle \hyperref[prin2]{\textbf{P2}} due to the lack of technical
measures and safeguards efficiently  enabling the exercise of the revocation right by websites.






\section{Recommendations}
\label{sec:disc}

In this section, we also propose recommendations to regulators for enhancing the implementation of consent revocation and  ensuring legal compliance, in the light of our findings.
\\

\noindent \textbf{R1. Need to unify EU requirements for consent revocation interface.} 
Our research shows that websites implement revocation 
inconsistently across the Web (see Table~\ref{tab:data_analysis1}). For example, the text placed in the website footer  differs across websites (e.g., ``Cookie Settings'', ``Privacy Settings'', but also ``EU Privacy''), and in many cases does not mention revocation. It is even harder to locate revocation links within privacy policies that are labelled with text such as ``How do I control cookies and how my data is used?'', or ``Managing our analytics cookies''. The vague and ambiguous representation of these options misleads the users and prevents them from exercising the revocation right  which should be clearly and distinctly recognisable (see \hyperref[lr1]{\textbf{LR1}}).
Icons, that are supported by regulators and are easy to locate, use different visualisations, which may also confuse the user.
EU DPAs should propose  \emph{unified interface requirements describing the interface, location, and wordings of revocation}. An inspiration can be taken from CPRA regulation in California, where 
three acceptable wordings are proposed, which simplifies locating it with automated means to measure compliance~\cite{Tran-etal-24-CHI}.\\

\noindent\textbf{R2. Need to standardize consent storage.}
We observed various inconsistencies in storing the {\tcss} and OneTrust consent strings (see \S\ref{sec:consistency}). This is due to the usage of different storage options that are not updated simultaneously when consent is revoked.
We \emph{invite regulators to establish standards for consent storage, and  implement security measures to protect  the integrity of consent strings}. \\

\noindent\textbf{R3. Need to standardize consent communication via event listeners.}
Our results show that only 43 out of 163 websites had third-parties installing event listeners in order to be updated about the change in consent decisions (see \S\ref{sec:informing-consent}). While OneTrust suggests developers to add event listeners to synchronize consent updates when integrating advertising systems~\cite{OT_GTM_integration}, 
our results show that IAB Europe TCF does not standardize event listeners, prompting every CMP to propose their own solution. 
This makes it harder for third-parties to adapt to different implementations across websites, placing the burden on them along with legal consequences (for continuous processing of personal data), without offering a practical solution to ensure compliance. 
We therefore recommend standard-setting bodies like IAB Europe TCF to standardize the implementation of event listeners and the callback functions returned by  event listeners to inform third-parties about consent revocation. 
We also propose \emph{regulators to take a position on the means of communicating consent to third-parties}. We believe that event listeners are able to provide reasonable means for communicating  revocation decisions, and help in the allocation of responsibility to collect valid consent by third-parties present on a website.\\ 

\noindent\textbf{R4. Need to regulate consent communication via HTTP requests.}
The \emph{most prevalent violation} we detected indicates that 74.2\% of websites do not properly inform all third-parties about consent revocation when the HTTP request method is used (see \S\ref{sec:informing-consent}). 
Our result shows that this concrete way of informing third-parties exempts  big advertising actors, such as \texttt{doubleclick.net} and \texttt{criteo.com} (see Figure \ref{fig:tp2-full}) from any responsibility regarding data deletion 
when users revoke consent.
Moreover, these third-parties  only receive  positive consent from the majority of websites and consequently, while illegally processing  data from users who revoked consent, these companies can falsely demonstrate  evidence of compliance to regulators by providing a \emph{false proof of consent}.
%
%
%
We therefore propose that \emph{EU regulators express their position regarding how websites should inform third-parties}, and whether  HTTP requests is an acceptable method, since our findings demonstrate that this approach leads to the most prevalent violations. 

\section{Conclusion}
\label{sec:conclusion}
%
In this work, we propose a framework to audit compliance of consent revocation on the Web. 
We found multiple instances of violations of the EU Data Protection law while analysing interfaces of revocation, including use of different interfaces (19.87\%) and more effort to revoke than to accept (20.5\%).
%
%
Multiple violations were observed on the usage of cookies and storing positive consent despite user's revocation. 
%
Most shockingly,  on 74\% of websites, third-parties that have received consent upon user's acceptance, are not informed of user's revocation, leading to the illegal processing of users' data by such third-parties. 
Our findings emphasise the need for improved legal compliance of consent revocation, in particular, proper, consistent, and uniform implementation of revocation communication and data deletion practices.


\section*{Acknowledgements}
This work has been supported by the ANR 22-PECY-0002 IPoP (Interdisciplinary Project on Privacy) project of the Cybersecurity PEPR, the TULIP project of the ANR MRSEI program 2023,  and the Inria International Chair funding.

%

\bibliographystyle{ACM-Reference-Format}
\bibliography{mybibliography}


\begin{thebibliography}{75}


\ifx \showCODEN    \undefined \def \showCODEN     #1{\unskip}     \fi
\ifx \showDOI      \undefined \def \showDOI       #1{#1}\fi
\ifx \showISBNx    \undefined \def \showISBNx     #1{\unskip}     \fi
\ifx \showISBNxiii \undefined \def \showISBNxiii  #1{\unskip}     \fi
\ifx \showISSN     \undefined \def \showISSN      #1{\unskip}     \fi
\ifx \showLCCN     \undefined \def \showLCCN      #1{\unskip}     \fi
\ifx \shownote     \undefined \def \shownote      #1{#1}          \fi
\ifx \showarticletitle \undefined \def \showarticletitle #1{#1}   \fi
\ifx \showURL      \undefined \def \showURL       {\relax}        \fi
\providecommand\bibfield[2]{#2}
\providecommand\bibinfo[2]{#2}
\providecommand\natexlab[1]{#1}
\providecommand\showeprint[2][]{arXiv:#2}

\bibitem[Anonymous(2025)]%
        {sup-material}
\bibfield{author}{\bibinfo{person}{Anonymous}.} \bibinfo{year}{2025}\natexlab{}.
\newblock \bibinfo{title}{Supplementary Material for Measuring Compliance of Consent Revocation on the Web}.
\newblock \bibinfo{howpublished}{\url{https://anonymous.4open.science/r/Measuring-Compliance-of-Consent-Revocation-on-the-Web-E4CB}}.
\newblock


\bibitem[AU(2022)]%
        {consent_o_matic}
\bibfield{author}{\bibinfo{person}{CAVI AU}.} \bibinfo{year}{2022}\natexlab{}.
\newblock \bibinfo{title}{Consent-O-Matic}.
\newblock \bibinfo{howpublished}{\url{https://consentomatic.au.dk/}}.
\newblock


\bibitem[Aziz and Wilson(2024)]%
        {AW-24-popets}
\bibfield{author}{\bibinfo{person}{Muhammad Abu~Bakar Aziz} {and} \bibinfo{person}{Christo Wilson}.} \bibinfo{year}{2024}\natexlab{}.
\newblock \showarticletitle{{Johnny Still Can’t Opt-out: Assessing the IAB CCPA Compliance Framework}}.
\newblock \bibinfo{journal}{\emph{Proc. Priv. Enhancing Technol.}} \bibinfo{volume}{2024}, \bibinfo{number}{4} (\bibinfo{year}{2024}), \bibinfo{pages}{349--363}.
\newblock
\urldef\tempurl%
\url{https://doi.org/10.56553/popets-2024-0120}
\showDOI{\tempurl}


\bibitem[{Bielova} et~al\mbox{.}(2024)]%
        {Biel-etal-24-JOLT}
\bibfield{author}{\bibinfo{person}{Nataliia {Bielova}}, \bibinfo{person}{Cristiana Santos}, {and} \bibinfo{person}{Colin~M Gray}.} \bibinfo{year}{2024}\natexlab{}.
\newblock \showarticletitle{Two worlds apart! {C}losing the gap between regulating {EU} consent and user studies}.
\newblock \bibinfo{journal}{\emph{Harvard Journal of Law \& Technology (JOLT)}}  \bibinfo{volume}{37} (\bibinfo{year}{2024}).
\newblock


\bibitem[Board(2007)]%
        {EDPB-4-07}
\bibfield{author}{\bibinfo{person}{European Data~Protection Board}.} \bibinfo{year}{2007}\natexlab{}.
\newblock \bibinfo{title}{Opinion 4/2007 on the concept of personal data ({WP} 136), adopted on 20.06.2007}.
\newblock
\newblock
\urldef\tempurl%
\url{https://ec.europa.eu/justice/article-29/documentation/opinion recommendation/files/2007/wp136_en.pdf}
\showURL{%
\tempurl}


\bibitem[Bollinger et~al\mbox{.}(2022)]%
        {automating_gdpr_violation_detection}
\bibfield{author}{\bibinfo{person}{Dino Bollinger}, \bibinfo{person}{Karel Kubicek}, \bibinfo{person}{Carlos Cotrini}, {and} \bibinfo{person}{David Basin}.} \bibinfo{year}{2022}\natexlab{}.
\newblock \showarticletitle{Automating Cookie Consent and {GDPR} Violation Detection}. In \bibinfo{booktitle}{\emph{31st USENIX Security Symposium}}. \bibinfo{pages}{2893--2910}.
\newblock


\bibitem[Bouhoula et~al\mbox{.}(2024)]%
        {Bouh-etal-24-USENIX}
\bibfield{author}{\bibinfo{person}{Ahmed Bouhoula}, \bibinfo{person}{Karel Kubicek}, \bibinfo{person}{Amit Zac}, \bibinfo{person}{Carlos Cotrini}, {and} \bibinfo{person}{David Basin}.} \bibinfo{year}{2024}\natexlab{}.
\newblock \showarticletitle{Automated, Large-Scale Analysis of Cookie Notice Compliance}. In \bibinfo{booktitle}{\emph{USENIX Security Symposium}}.
\newblock


\bibitem[{California State Legislature}(2018)]%
        {ccpa}
\bibfield{author}{\bibinfo{person}{{California State Legislature}}.} \bibinfo{year}{2018}\natexlab{}.
\newblock \bibinfo{title}{California Consumer Privacy Act of 2018}.
\newblock
\newblock
\urldef\tempurl%
\url{https://oag.ca.gov/privacy/ccpa}
\showURL{%
\tempurl}


\bibitem[CJEU-C‑129/21-2022(2022)]%
        {CJEU-C‑129/21-2022}
CJEU-C‑129/21-2022 \bibinfo{year}{2022}\natexlab{}.
\newblock \bibinfo{title}{Judgment in Case C-129/21 Proximus NV v Gegevensbeschermingsautoriteit}.
\newblock
\newblock
\urldef\tempurl%
\url{{https://eur-lex.europa.eu/legal-content/EN/TXT/?uri=CELEX%3A62021CJ0129}}
\showURL{%
\tempurl}


\bibitem[CNIL(2023)]%
        {CNIL-revokeDecision}
\bibfield{author}{\bibinfo{person}{CNIL}.} \bibinfo{year}{2023}\natexlab{}.
\newblock \bibinfo{title}{Délibération de la formation restreinte n°SAN-2023-024 du 29 décembre 2023 concernant la société YAHOO EMEA LIMITED}.
\newblock \bibinfo{howpublished}{\url{https://www.legifrance.gouv.fr/cnil/id/CNILTEXT000048967251}}.
\newblock


\bibitem[{Commission Nationale de l'Informatique et des Libertés (CNIL)}(2020)]%
        {CNIL-recoms-2020}
\bibfield{author}{\bibinfo{person}{{Commission Nationale de l'Informatique et des Libertés (CNIL)}}.} \bibinfo{year}{2020}\natexlab{}.
\newblock \bibinfo{title}{Recommandation ``cookies et autres traceurs''}.
\newblock
\newblock
\urldef\tempurl%
\url{https://www.cnil.fr/sites/default/files/atoms/files/recommandation-cookies-et-autres-traceurs.pdf}
\showURL{%
\tempurl}


\bibitem[{Danish DPA (Datatilsynet)}(2021)]%
        {Danish-guidelines-2021}
\bibfield{author}{\bibinfo{person}{{Danish DPA (Datatilsynet)}}.} \bibinfo{year}{2021}\natexlab{}.
\newblock \bibinfo{title}{Quick Guide on the use of cookies}.
\newblock
\newblock
\urldef\tempurl%
\url{{https://www.datatilsynet.dk/Media/F/8/Behandling\%20af\%20personoplysninger\%20om\%20hjemmesidebes\%C3\%B8gende.pdf}}
\showURL{%
\tempurl}


\bibitem[{Data Protection Commission (DPC)}(2020)]%
        {Irish-guidelines-2020}
\bibfield{author}{\bibinfo{person}{{Data Protection Commission (DPC)}}.} \bibinfo{year}{2020}\natexlab{}.
\newblock \bibinfo{title}{Guidance Note: Cookies and other Tracking Technologies}.
\newblock
\newblock
\urldef\tempurl%
\url{{https://www.dataprotection.ie/sites/default/files/uploads/2020-04/Guidance\%20note\%20on\%20cookies\%20and\%20other\%20tracking\%20technologies.pdf}}
\showURL{%
\tempurl}


\bibitem[Degeling et~al\mbox{.}(2019)]%
        {ndss19}
\bibfield{author}{\bibinfo{person}{Martin Degeling}, \bibinfo{person}{Christine Utz}, \bibinfo{person}{Christopher Lentzsch}, \bibinfo{person}{Henry Hosseini}, \bibinfo{person}{Florian Schaub}, {and} \bibinfo{person}{Thorsten Holz}.} \bibinfo{year}{2019}\natexlab{}.
\newblock \showarticletitle{{We Value Your Privacy... Now take some cookies: Measuring the GDPR's impact on web privacy}}. In \bibinfo{booktitle}{\emph{Proceedings of the 26th Network and Distributed System Security Symposium}}.
\newblock


\bibitem[{Directorate General Justice, European Commission}(2013)]%
        {wp208}
\bibfield{author}{\bibinfo{person}{{Directorate General Justice, European Commission}}.} \bibinfo{year}{2013}\natexlab{}.
\newblock \bibinfo{title}{Working Document 02/2013 providing guidance on obtaining consent for cookies}.
\newblock
\newblock
\urldef\tempurl%
\url{https://ec.europa.eu/justice/article-29/documentation/opinion-recommendation/files/2013/wp208_en.pdf}
\showURL{%
\tempurl}
\newblock
\shownote{[Online; accessed 2025-02-25]}.


\bibitem[DPA(2024)]%
        {Garante-revokeDecision}
\bibfield{author}{\bibinfo{person}{Italian DPA}.} \bibinfo{year}{2024}\natexlab{}.
\newblock \bibinfo{title}{Provvedimento del 6 giugno 2024 [10029424] against Eni Plenitude S.p.A.}
\newblock \bibinfo{howpublished}{\url{https://www.garanteprivacy.it/web/guest/home/docweb/-/docweb-display/docweb/10029424}}.
\newblock


\bibitem[DSK-DPA-cookies-2021(2021)]%
        {DSK-DPA-cookies-2021}
DSK-DPA-cookies-2021 \bibinfo{year}{2021}\natexlab{}.
\newblock \bibinfo{title}{{Guidance from the Conference of Independent Data Protection Supervisory Authorities of the Federal Government and the States of 20 December 2021 (OH Telemedia 2021, V.1.1)}}.
\newblock
\newblock
\newblock
\shownote{\url{https://www.datenschutzkonferenz-online.de/media/oh/20211220_oh_telemedien.pdf}, accessed on 2024.09.03}.


\bibitem[Du et~al\mbox{.}(2024)]%
        {mowchecker}
\bibfield{author}{\bibinfo{person}{Xiaolin Du}, \bibinfo{person}{Zhemin Yang}, \bibinfo{person}{Jiapeng Lin}, \bibinfo{person}{Yinzhi Cao}, {and} \bibinfo{person}{Min Yang}.} \bibinfo{year}{2024}\natexlab{}.
\newblock \showarticletitle{{ Withdrawing is believing? Detecting Inconsistencies between Withdrawal Choices and Third-party Data Collections in Mobile Apps }}. In \bibinfo{booktitle}{\emph{2024 IEEE Symposium on Security and Privacy (SP)}}. \bibinfo{pages}{735--751}.
\newblock


\bibitem[DutchDPA-revocationDoc-2024(2024)]%
        {DutchDPA-revocationDoc-2024}
DutchDPA-revocationDoc-2024 \bibinfo{year}{2024}\natexlab{}.
\newblock \bibinfo{title}{{Explanation of the standard of the AP on the withdrawal of consent for cookie banners 01 March 2024}}.
\newblock
\newblock
\newblock
\shownote{\url{https://www.autoriteitpersoonsgegevens.nl/documenten/normuitleg-ap-over-intrekken-van-toestemming-bij-cookiebanners}, accessed on 2024.09.03}.


\bibitem[(EDPB)(2013)]%
        {EDPB-PurposeLimitation}
\bibfield{author}{\bibinfo{person}{European Data Protection~Board (EDPB)}.} \bibinfo{year}{2013}\natexlab{}.
\newblock \bibinfo{title}{Opinion 03/2013 on purpose limitation (WP 203)}.
\newblock \bibinfo{howpublished}{Available at \url{https://ec.europa.eu/justice/article-29/documentation/opinion-recommendation/files/2013/wp203_en.pdf}}.
\newblock


\bibitem[(EDPB)(2019)]%
        {EDPB-guidelinesGDPR-ePD}
\bibfield{author}{\bibinfo{person}{European Data Protection~Board (EDPB)}.} \bibinfo{year}{2019}\natexlab{}.
\newblock \bibinfo{title}{Opinion 5/2019 on the Interplay between the ePrivacy Directive and the GDPR, in Particular Regarding the Competence, Tasks and Powers of Data Protection Authorities}.
\newblock
\newblock
\urldef\tempurl%
\url{https://edpb.europa.eu/sites/edpb/files/files/file1/201905_edpb_opinion_eprivacydir_gdpr_interplay_en_0.pdf}
\showURL{%
\tempurl}


\bibitem[ePD-09(2009)]%
        {ePD-09}
ePD-09 \bibinfo{year}{2009}\natexlab{}.
\newblock \bibinfo{title}{{Directive 2009/136/EC of the European Parliament and of the Council of 25 November 2009}}.
\newblock
\newblock
\newblock
\shownote{\url{https://eur-lex.europa.eu/legal-content/EN/TXT/?uri=celex\%3A32009L0136}, accessed on 2019.10.31}.


\bibitem[Europe(2025)]%
        {TCF}
\bibfield{author}{\bibinfo{person}{IAB Europe}.} \bibinfo{year}{2025}\natexlab{}.
\newblock \bibinfo{title}{TCF - Transparency \& Consent Framework - IAB Europe}.
\newblock \bibinfo{howpublished}{\url{https://iabeurope.eu/transparency-consent-framework/}}.
\newblock


\bibitem[{European Data Protection Board}(2020a)]%
        {EDPB-05-2020consent}
\bibfield{author}{\bibinfo{person}{{European Data Protection Board}}.} \bibinfo{year}{2020}\natexlab{a}.
\newblock \bibinfo{title}{Guidelines 05/2020 on consent under Regulation 2016/679}.
\newblock
\newblock
\newblock
\shownote{\url{https://edpb.europa.eu/sites/edpb/files/files/file1/edpb_guidelines_202005_consent_en.pdf}}.


\bibitem[{European Data Protection Board}(2020b)]%
        {EDPB2020-DPbDbDesign}
\bibfield{author}{\bibinfo{person}{{European Data Protection Board}}.} \bibinfo{year}{2020}\natexlab{b}.
\newblock \bibinfo{title}{Guidelines 4/2019 on Article 25 Data Protection by Design and by Default Version 2.0 Adopted on 20 October 2020}.
\newblock
\newblock
\urldef\tempurl%
\url{https://www.edpb.europa.eu/sites/default/files/files/file1/edpb_guidelines_201904_dataprotection_by_design_and_by_default_v2.0_en.pdf}
\showURL{%
\tempurl}


\bibitem[{European Data Protection Board}(2022)]%
        {EDPB2022-Darkpatterns}
\bibfield{author}{\bibinfo{person}{{European Data Protection Board}}.} \bibinfo{year}{2022}\natexlab{}.
\newblock \bibinfo{title}{Guidelines 3/2022 on Dark patterns in social media platform interfaces: How to recognise and avoid them}.
\newblock
\newblock
\urldef\tempurl%
\url{https://edpb.europa.eu/our-work-tools/documents/public-consultations/2022/guidelines-32022-dark-patterns-social-media_en}
\showURL{%
\tempurl}


\bibitem[{European Data Protection Board}(2023)]%
        {EDPB-tf-2023}
\bibfield{author}{\bibinfo{person}{{European Data Protection Board}}.} \bibinfo{year}{2023}\natexlab{}.
\newblock \bibinfo{title}{Report of the work undertaken by the Cookie Banner Taskforce}.
\newblock
\newblock
\newblock
\shownote{\url{https://edpb.europa.eu/our-work-tools/our-documents/report/report-work-undertaken-cookie-banner-taskforce_en}}.


\bibitem[{European Data Protection Board}(2024)]%
        {EDPB2024-CookiePaywalls}
\bibfield{author}{\bibinfo{person}{{European Data Protection Board}}.} \bibinfo{year}{2024}\natexlab{}.
\newblock \bibinfo{title}{Opinion 08/2024 on Valid Consent in the Context of Consent or Pay Models Implemented by Large Online Platforms, Adopted on 17 April 2024}.
\newblock
\newblock
\newblock
\shownote{\url{https://www.edpb.europa.eu/system/files/2024-04/edpb_opinion_202408_consentorpay_en.pdf}}.


\bibitem[{European Data Protection Board (EDPB), Article 29 Working Party}(2012)]%
        {EDPB-Cookie-Exemption}
\bibfield{author}{\bibinfo{person}{{European Data Protection Board (EDPB), Article 29 Working Party}}.} \bibinfo{year}{2012}\natexlab{}.
\newblock \bibinfo{title}{{Opinion 04/2012 on Cookie Consent Exemption ({WP} 194)}}.
\newblock
\newblock


\bibitem[{European Parliament} and {Council of the European Union}(2016)]%
        {gdpr}
\bibfield{author}{\bibinfo{person}{{European Parliament}} {and} \bibinfo{person}{{Council of the European Union}}.} \bibinfo{year}{2016}\natexlab{}.
\newblock \bibinfo{title}{Regulation ({EU}) 2016/679 of the {European} {Parliament} and of the {Council}}.
\newblock
\newblock
\urldef\tempurl%
\url{https://data.europa.eu/eli/reg/2016/679/oj}
\showURL{%
\tempurl}


\bibitem[Fanboy and Khrin(2023)]%
        {easylist}
\bibfield{author}{\bibinfo{person}{Famlam Fanboy, MonztA} {and} \bibinfo{person}{Khrin}.} \bibinfo{year}{2023}\natexlab{}.
\newblock \bibinfo{title}{EasyList}.
\newblock \bibinfo{howpublished}{\url{https://easylist.to/}}.
\newblock
\newblock
\shownote{Accessed on 12 March 2023}.


\bibitem[for Digital~Government(2023)]%
        {DanishAgency-revokeDecision}
\bibfield{author}{\bibinfo{person}{Danish~Agency for Digital~Government}.} \bibinfo{year}{2023}\natexlab{}.
\newblock \bibinfo{title}{Danish Agency for Digital Government Decision against Meta}.
\newblock \bibinfo{howpublished}{\url{https://gdprhub.eu/images/b/bb/Paabud-til-meta-platforms.pdf }}.
\newblock


\bibitem[Fouad et~al\mbox{.}(2020)]%
        {Foua-etal-20-IWPE}
\bibfield{author}{\bibinfo{person}{Imane Fouad}, \bibinfo{person}{Cristiana Santos}, \bibinfo{person}{Feras Al~Kassar}, \bibinfo{person}{Nataliia Bielova}, {and} \bibinfo{person}{Stefano Calzavara}.} \bibinfo{year}{2020}\natexlab{}.
\newblock \showarticletitle{On Compliance of Cookie Purposes with the Purpose Specification Principle}. In \bibinfo{booktitle}{\emph{2020 IEEE European Symposium on Security and Privacy Workshops (EuroS\&PW)}}. \bibinfo{pages}{326--333}.
\newblock
\urldef\tempurl%
\url{https://doi.org/10.1109/EuroSPW51379.2020.00051}
\showDOI{\tempurl}


\bibitem[Giese and Stabauer(2022)]%
        {10.1007/978-3-031-05544-7_21}
\bibfield{author}{\bibinfo{person}{Julia Giese} {and} \bibinfo{person}{Martin Stabauer}.} \bibinfo{year}{2022}\natexlab{}.
\newblock \showarticletitle{Factors That Influence Cookie Acceptance: Characteristics of Cookie Notices That Users Perceive to Affect Their Decisions}. In \bibinfo{booktitle}{\emph{9th International Conference on HCI in Business, Government and Organizations}}. \bibinfo{pages}{272–285}.
\newblock
\showISBNx{978-3-031-05543-0}


\bibitem[Gray et~al\mbox{.}(2021)]%
        {dark_pattern_legal_req}
\bibfield{author}{\bibinfo{person}{Colin~M. Gray}, \bibinfo{person}{Cristiana Santos}, \bibinfo{person}{Nataliia Bielova}, \bibinfo{person}{Michael Toth}, {and} \bibinfo{person}{Damian Clifford}.} \bibinfo{year}{2021}\natexlab{}.
\newblock \showarticletitle{Dark Patterns and the Legal Requirements of Consent Banners: An Interaction Criticism Perspective}. In \bibinfo{booktitle}{\emph{Proceedings of the 2021 CHI Conference on Human Factors in Computing Systems}}. Article \bibinfo{articleno}{172}, \bibinfo{numpages}{18}~pages.
\newblock


\bibitem[Habib et~al\mbox{.}(2022)]%
        {Habib-etal-2022}
\bibfield{author}{\bibinfo{person}{Hana Habib}, \bibinfo{person}{Megan Li}, \bibinfo{person}{Ellie Young}, {and} \bibinfo{person}{Lorrie Cranor}.} \bibinfo{year}{2022}\natexlab{}.
\newblock \showarticletitle{{“Okay, Whatever”: An Evaluation of Cookie Consent Interfaces}}. In \bibinfo{booktitle}{\emph{Proceedings of the 2022 CHI Conference on Human Factors in Computing Systems}}. Article \bibinfo{articleno}{621}, \bibinfo{numpages}{27}~pages.
\newblock
\showISBNx{9781450391573}


\bibitem[Habib et~al\mbox{.}(2019)]%
        {data-deletion-opt-out-choices}
\bibfield{author}{\bibinfo{person}{Hana Habib}, \bibinfo{person}{Yixin Zou}, \bibinfo{person}{Aditi Jannu}, \bibinfo{person}{Neha Sridhar}, \bibinfo{person}{Chelse Swoopes}, \bibinfo{person}{Alessandro Acquisti}, \bibinfo{person}{Lorrie~Faith Cranor}, \bibinfo{person}{Norman Sadeh}, {and} \bibinfo{person}{Florian Schaub}.} \bibinfo{year}{2019}\natexlab{}.
\newblock \showarticletitle{An Empirical Analysis of Data Deletion and {Opt-Out} Choices on 150 Websites}. In \bibinfo{booktitle}{\emph{Fifteenth Symposium on Usable Privacy and Security (SOUPS 2019)}}. \bibinfo{pages}{387--406}.
\newblock
\showISBNx{978-1-939133-05-2}


\bibitem[Hils et~al\mbox{.}(2020)]%
        {imc20}
\bibfield{author}{\bibinfo{person}{Maximilian Hils}, \bibinfo{person}{Daniel~W. Woods}, {and} \bibinfo{person}{Rainer B\"{o}hme}.} \bibinfo{year}{2020}\natexlab{}.
\newblock \showarticletitle{Measuring the Emergence of Consent Management on the Web}. In \bibinfo{booktitle}{\emph{Proceedings of the ACM Internet Measurement Conference}} \emph{(\bibinfo{series}{IMC '20})}. \bibinfo{pages}{317–332}.
\newblock
\showISBNx{9781450381383}


\bibitem[Hils et~al\mbox{.}(2021)]%
        {hills-2021-pets}
\bibfield{author}{\bibinfo{person}{Maximilian Hils}, \bibinfo{person}{Daniel~W. Woods}, {and} \bibinfo{person}{Rainer B{\"o}hme}.} \bibinfo{year}{2021}\natexlab{}.
\newblock \showarticletitle{Privacy Preference Signals: {P}ast, Present and Future}.
\newblock \bibinfo{journal}{\emph{Proceedings on Privacy Enhancing Technologies}} \bibinfo{number}{4} (\bibinfo{year}{2021}).
\newblock
\urldef\tempurl%
\url{https://petsymposium.org/2021/files/papers/issue4/popets-2021-0069.pdf}
\showURL{%
\tempurl}


\bibitem[I dont care about cookies({[n.\,d.]})]%
        {i_dont_care_about_cookies}
I dont care about cookies \bibinfo{year}{[n.\,d.]}\natexlab{}.
\newblock \bibinfo{title}{I don't care about cookies 3.5.0}.
\newblock \bibinfo{howpublished}{\url{https://www.i-dont-care-about-cookies.eu/}}.
\newblock


\bibitem[IAB(2024)]%
        {TCF_consentstring}
\bibfield{author}{\bibinfo{person}{IAB}.} \bibinfo{year}{2024}\natexlab{}.
\newblock \bibinfo{title}{GDPR-Transparency-and-Consent-Framework/TCFv2/IAB Tech Lab - Consent string and vendor list formats v2.md at master - InteractiveAdvertisingBureau/GDPR-Transparency-and-Consent-Framework}.
\newblock \bibinfo{howpublished}{\url{https://github.com/InteractiveAdvertisingBureau/GDPR-Transparency-and-Consent-Framework/blob/master/TCFv2/}}.
\newblock


\bibitem[IAB(2025a)]%
        {TCF_purposes}
\bibfield{author}{\bibinfo{person}{IAB}.} \bibinfo{year}{2025}\natexlab{a}.
\newblock \bibinfo{title}{Index - Global Vendor List}.
\newblock \bibinfo{howpublished}{\url{https://register.consensu.org/Translation}}.
\newblock


\bibitem[IAB(2025b)]%
        {TCFv22}
\bibfield{author}{\bibinfo{person}{IAB}.} \bibinfo{year}{2025}\natexlab{b}.
\newblock \bibinfo{title}{TCF Policies - TransparencyConsentFramework\_Policies\_Version 2024-06-3.5.0.docx}.
\newblock \bibinfo{howpublished}{\url{https://iabeurope.eu/transparency-consent-framework-file/TCF\%20Policies\%20-\%20TransparencyConsentFramework_Policies_Version\%202024-06-3.5.0.pdf}}.
\newblock


\bibitem[{Information Commissioner's Office (ICO)}(2020)]%
        {UKDPA-callCP-2024}
\bibfield{author}{\bibinfo{person}{{Information Commissioner's Office (ICO)}}.} \bibinfo{year}{2020}\natexlab{}.
\newblock \bibinfo{title}{Call for views on "consent or pay" business models}.
\newblock
\newblock
\urldef\tempurl%
\url{{https://ico.org.uk/cookies-call-for-views-202403}}
\showURL{%
\tempurl}


\bibitem[Jesus and Pandit(2022)]%
        {484c16b527a24fc9b5df872de42eb1ec}
\bibfield{author}{\bibinfo{person}{Vitor Jesus} {and} \bibinfo{person}{Harshvardhan~J. Pandit}.} \bibinfo{year}{2022}\natexlab{}.
\newblock \showarticletitle{Consent Receipts for a Usable and Auditable Web of Personal Data}.
\newblock \bibinfo{journal}{\emph{IEEE Access}}  \bibinfo{volume}{10} (\bibinfo{year}{2022}), \bibinfo{pages}{28545--28563}.
\newblock
\urldef\tempurl%
\url{https://doi.org/10.1109/ACCESS.2022.3157850}
\showDOI{\tempurl}


\bibitem[Kampanos and Shahandashti(2021)]%
        {acceptall}
\bibfield{author}{\bibinfo{person}{Georgios Kampanos} {and} \bibinfo{person}{Siamak~F. Shahandashti}.} \bibinfo{year}{2021}\natexlab{}.
\newblock \showarticletitle{Accept All: The Landscape of Cookie Banners in Greece and the UK}. In \bibinfo{booktitle}{\emph{ICT Systems Security and Privacy Protection}}. \bibinfo{pages}{213--227}.
\newblock
\showISBNx{978-3-030-78120-0}


\bibitem[Khandelwal et~al\mbox{.}(2021)]%
        {prisec}
\bibfield{author}{\bibinfo{person}{Rishabh Khandelwal}, \bibinfo{person}{Thomas Linden}, \bibinfo{person}{Hamza Harkous}, {and} \bibinfo{person}{Kassem Fawaz}.} \bibinfo{year}{2021}\natexlab{}.
\newblock \showarticletitle{{PriSEC: A Privacy Settings Enforcement Controller}}. In \bibinfo{booktitle}{\emph{30th USENIX Security Symposium (USENIX Security 21)}}. \bibinfo{pages}{465--482}.
\newblock


\bibitem[Khandelwal et~al\mbox{.}(2023)]%
        {cookie-enforcer}
\bibfield{author}{\bibinfo{person}{Rishabh Khandelwal}, \bibinfo{person}{Asmit Nayak}, \bibinfo{person}{Hamza Harkous}, {and} \bibinfo{person}{Kassem Fawaz}.} \bibinfo{year}{2023}\natexlab{}.
\newblock \showarticletitle{Automated cookie notice analysis and enforcement}. In \bibinfo{booktitle}{\emph{Proceedings of the 32nd USENIX Conference on Security Symposium}}. Article \bibinfo{articleno}{63}, \bibinfo{numpages}{18}~pages.
\newblock
\showISBNx{978-1-939133-37-3}


\bibitem[Kirkman et~al\mbox{.}(2023)]%
        {darkdialogs}
\bibfield{author}{\bibinfo{person}{D. Kirkman}, \bibinfo{person}{K. Vaniea}, {and} \bibinfo{person}{D.~W. Woods}.} \bibinfo{year}{2023}\natexlab{}.
\newblock \showarticletitle{{DarkDialogs: Automated Detection of 10 Dark Patterns on Cookie Dialogs}}. In \bibinfo{booktitle}{\emph{2023 IEEE 8th European Symposium on Security and Privacy (EuroSP)}}. \bibinfo{pages}{847--867}.
\newblock


\bibitem[Kulyk et~al\mbox{.}(2018)]%
        {usec2018}
\bibfield{author}{\bibinfo{person}{Oksana Kulyk}, \bibinfo{person}{Annika Hilt}, \bibinfo{person}{Nina Gerber}, {and} \bibinfo{person}{Melanie Volkamer}.} \bibinfo{year}{2018}\natexlab{}.
\newblock \showarticletitle{{"This Website Uses Cookies": Users' Perceptions and Reactions to the Cookie Disclaimer}}. In \bibinfo{booktitle}{\emph{3rd European Workshop on Usable Security}}.
\newblock


\bibitem[Lab and Europe(2018)]%
        {tcf11-url}
\bibfield{author}{\bibinfo{person}{IAB~Tech Lab} {and} \bibinfo{person}{IAB Europe}.} \bibinfo{year}{2018}\natexlab{}.
\newblock \bibinfo{title}{GDPR consent passing for URL-based services: Transparency and consent framework}.
\newblock
\newblock
\newblock
\shownote{\url{https://github.com/InteractiveAdvertisingBureau/ GDPR-Transparency-and-Consent-Framework/blob/master}}.


\bibitem[{Le Pochat} et~al\mbox{.}(2019)]%
        {tranco}
\bibfield{author}{\bibinfo{person}{Victor {Le Pochat}}, \bibinfo{person}{Tom {Van Goethem}}, \bibinfo{person}{Samaneh Tajalizadehkhoob}, \bibinfo{person}{Maciej Korczy\'{n}ski}, {and} \bibinfo{person}{Wouter Joosen}.} \bibinfo{year}{2019}\natexlab{}.
\newblock \showarticletitle{Tranco: A Research-Oriented Top Sites Ranking Hardened Against Manipulation}. In \bibinfo{booktitle}{\emph{Proceedings of the 26th Network and Distributed System Security Symposium}}.
\newblock


\bibitem[Leenes and Kosta(2015)]%
        {taming-cookie-monster}
\bibfield{author}{\bibinfo{person}{Ronald Leenes} {and} \bibinfo{person}{Eleni Kosta}.} \bibinfo{year}{2015}\natexlab{}.
\newblock \showarticletitle{Taming the cookie monster with Dutch law – A tale of regulatory failure}.
\newblock \bibinfo{journal}{\emph{Computer Law \& Security Review}}  \bibinfo{volume}{31} (\bibinfo{date}{03} \bibinfo{year}{2015}).
\newblock
\urldef\tempurl%
\url{https://doi.org/10.1016/j.clsr.2015.01.004}
\showDOI{\tempurl}


\bibitem[Ma and Birrell(2022)]%
        {10.1145/3491101.3519687}
\bibfield{author}{\bibinfo{person}{Eryn Ma} {and} \bibinfo{person}{Eleanor Birrell}.} \bibinfo{year}{2022}\natexlab{}.
\newblock \showarticletitle{Prospective Consent: The Effect of Framing on Cookie Consent Decisions}. In \bibinfo{booktitle}{\emph{Extended Abstracts of the 2022 CHI Conference on Human Factors in Computing Systems}}. Article \bibinfo{articleno}{400}, \bibinfo{numpages}{6}~pages.
\newblock
\showISBNx{9781450391566}


\bibitem[Machuletz and Boehme(2020)]%
        {popets2020}
\bibfield{author}{\bibinfo{person}{Dominique Machuletz} {and} \bibinfo{person}{Rainer Boehme}.} \bibinfo{year}{2020}\natexlab{}.
\newblock \showarticletitle{{Multiple Purposes, Multiple Problems: A User Study of Consent Dialogs after GDPR}}. In \bibinfo{booktitle}{\emph{Proceedings on Privacy Enhancing Technologies Symposium}}, Vol.~\bibinfo{volume}{2}. \bibinfo{pages}{481--498}.
\newblock


\bibitem[Mathur et~al\mbox{.}(2021)]%
        {darkpattern}
\bibfield{author}{\bibinfo{person}{Arunesh Mathur}, \bibinfo{person}{Mihir Kshirsagar}, {and} \bibinfo{person}{Jonathan Mayer}.} \bibinfo{year}{2021}\natexlab{}.
\newblock \showarticletitle{What Makes a Dark Pattern... Dark? Design Attributes, Normative Considerations, and Measurement Methods}. In \bibinfo{booktitle}{\emph{Proceedings of the 2021 CHI Conference on Human Factors in Computing Systems}} (Yokohama, Japan) \emph{(\bibinfo{series}{CHI '21})}. Article \bibinfo{articleno}{360}, \bibinfo{numpages}{18}~pages.
\newblock
\showISBNx{9781450380966}


\bibitem[Matte et~al\mbox{.}(2020a)]%
        {do_cookie_banners_respect}
\bibfield{author}{\bibinfo{person}{C{\'{e}}lestin Matte}, \bibinfo{person}{Nataliia Bielova}, {and} \bibinfo{person}{Cristiana Santos}.} \bibinfo{year}{2020}\natexlab{a}.
\newblock \showarticletitle{{Do Cookie Banners Respect my Choice? Measuring Legal Compliance of Banners from IAB Europe's Transparency and Consent Framework}}. In \bibinfo{booktitle}{\emph{2020 IEEE Symposium on Security and Privacy (SP)}}, Vol.~\bibinfo{volume}{1}. \bibinfo{pages}{791--809}.
\newblock


\bibitem[Matte et~al\mbox{.}(2020b)]%
        {Matt-etal-20-APF}
\bibfield{author}{\bibinfo{person}{C{\'e}lestin Matte}, \bibinfo{person}{Cristiana Santos}, {and} \bibinfo{person}{Nataliia Bielova}.} \bibinfo{year}{2020}\natexlab{b}.
\newblock \showarticletitle{{Purposes in IAB Europe's TCF: which legal basis and how are they used by advertisers?}}. In \bibinfo{booktitle}{\emph{{APF 2020 - Annual Privacy Forum}}}. \bibinfo{address}{Lisbon, Portugal}, \bibinfo{pages}{1--24}.
\newblock
\urldef\tempurl%
\url{https://inria.hal.science/hal-02566891}
\showURL{%
\tempurl}


\bibitem[Ninja Cookie({[n.\,d.]})]%
        {ninja_cookies}
Ninja Cookie \bibinfo{year}{[n.\,d.]}\natexlab{}.
\newblock \bibinfo{title}{Download Ninja Cookie - MajorGeeks}.
\newblock \bibinfo{howpublished}{\url{https://www.majorgeeks.com/files/details/ninja_cookie.html}}.
\newblock


\bibitem[Nouwens et~al\mbox{.}(2020)]%
        {dark_patterns_after_gdpr}
\bibfield{author}{\bibinfo{person}{Midas Nouwens}, \bibinfo{person}{Ilaria Liccardi}, \bibinfo{person}{Michael Veale}, \bibinfo{person}{David~R. Karger}, {and} \bibinfo{person}{Lalana Kagal}.} \bibinfo{year}{2020}\natexlab{}.
\newblock \showarticletitle{Dark Patterns after the {GDPR:} Scraping Consent Pop-ups and Demonstrating their Influence}.
\newblock \bibinfo{journal}{\emph{CoRR}}  \bibinfo{volume}{abs/2001.02479} (\bibinfo{year}{2020}).
\newblock
\showeprint[arXiv]{2001.02479}
\urldef\tempurl%
\url{http://arxiv.org/abs/2001.02479}
\showURL{%
\tempurl}


\bibitem[NOYB(2024)]%
        {noyb-2024-darkpatterns}
\bibfield{author}{\bibinfo{person}{NOYB}.} \bibinfo{year}{2024}\natexlab{}.
\newblock \bibinfo{title}{Consent Banner Report - Overview of EU and national guidelines on dark patterns}.
\newblock \bibinfo{howpublished}{\url{https://noyb.eu/sites/default/files/2024-07/noyb_Cookie_Report_2024.pdf}}.
\newblock


\bibitem[OneTrust({[n.\,d.]})]%
        {onetrust-dev}
OneTrust \bibinfo{year}{[n.\,d.]}\natexlab{}.
\newblock \bibinfo{title}{OneTrust Developer Portal}.
\newblock \bibinfo{howpublished}{\url{https://developer.onetrust.com/}}.
\newblock


\bibitem[OneTrust(2024)]%
        {OT_GTM_integration}
\bibfield{author}{\bibinfo{person}{OneTrust}.} \bibinfo{year}{2024}\natexlab{}.
\newblock \bibinfo{title}{Cookie Consent Integration with Google Tag Manager | MyOneTrust}.
\newblock \bibinfo{howpublished}{\url{https://my.onetrust.com/s/article/UUID-301b21c8-a73a-05e8-175a-36c9036728dc?language=en_US}}.
\newblock


\bibitem[Pandit et~al\mbox{.}(2019)]%
        {pandit_2019}
\bibfield{author}{\bibinfo{person}{Harshvardhan Pandit}, \bibinfo{person}{Christophe Debruyne}, \bibinfo{person}{Declan O’Sullivan}, {and} \bibinfo{person}{David Lewis}.} \bibinfo{year}{2019}\natexlab{}.
\newblock \showarticletitle{{GConsent} - {A} Consent Ontology Based on the {GDPR}}. In \bibinfo{booktitle}{\emph{The Semantic Web (ESWC 2019)}}. \bibinfo{publisher}{Springer}, \bibinfo{pages}{270--282}.
\newblock
\showISBNx{978-3-030-21347-3}
\urldef\tempurl%
\url{https://doi.org/10.1007/978-3-030-21348-0_18}
\showDOI{\tempurl}


\bibitem[Sanchez-Rola et~al\mbox{.}(2019)]%
        {asiaccs19}
\bibfield{author}{\bibinfo{person}{Iskander Sanchez-Rola}, \bibinfo{person}{Matteo Dell'Amico}, \bibinfo{person}{Platon Kotzias}, \bibinfo{person}{Davide Balzarotti}, \bibinfo{person}{Leyla Bilge}, \bibinfo{person}{Pierre-Antoine Vervier}, {and} \bibinfo{person}{Igor Santos}.} \bibinfo{year}{2019}\natexlab{}.
\newblock \showarticletitle{Can I Opt Out Yet? GDPR and the Global Illusion of Cookie Control}. In \bibinfo{booktitle}{\emph{Proceedings of the 2019 ACM Asia Conference on Computer and Communications Security}} (Auckland, New Zealand) \emph{(\bibinfo{series}{AsiaCCS '19})}. \bibinfo{pages}{340–351}.
\newblock
\showISBNx{9781450367523}


\bibitem[Santos et~al\mbox{.}(2020)]%
        {Sant_etal_20_TechReg}
\bibfield{author}{\bibinfo{person}{Cristiana Santos}, \bibinfo{person}{Nataliia {Bielova}}, {and} \bibinfo{person}{C{é}lestin Matte}.} \bibinfo{year}{2020}\natexlab{}.
\newblock \showarticletitle{Are cookie banners indeed compliant with the law? {D}eciphering {EU} legal requirements on consent and technical means to verify compliance of cookie banners}.
\newblock \bibinfo{journal}{\emph{Technology and Regulation (TechReg)}} (\bibinfo{year}{2020}), \bibinfo{pages}{91--135}.
\newblock
\newblock
\shownote{\url{https://doi.org/10.26116/techreg.2020.009}}.


\bibitem[Smith et~al\mbox{.}(2024)]%
        {Smith-WWW-24}
\bibfield{author}{\bibinfo{person}{Michael Smith}, \bibinfo{person}{Antonio Torres-Ag\"{u}ero}, \bibinfo{person}{Riley Grossman}, \bibinfo{person}{Pritam Sen}, \bibinfo{person}{Yi Chen}, {and} \bibinfo{person}{Cristian Borcea}.} \bibinfo{year}{2024}\natexlab{}.
\newblock \showarticletitle{A Study of GDPR Compliance under the Transparency and Consent Framework}. In \bibinfo{booktitle}{\emph{Proceedings of the ACM on Web Conference 2024}} (Singapore, Singapore). \bibinfo{pages}{1227–1236}.
\newblock
\showISBNx{9798400701719}


\bibitem[Soe et~al\mbox{.}(2020)]%
        {nordichi20}
\bibfield{author}{\bibinfo{person}{Than~Htut Soe}, \bibinfo{person}{Oda~Elise Nordberg}, \bibinfo{person}{Frode Guribye}, {and} \bibinfo{person}{Marija Slavkovik}.} \bibinfo{year}{2020}\natexlab{}.
\newblock \showarticletitle{Circumvention by Design - Dark Patterns in Cookie Consent for Online News Outlets}. In \bibinfo{booktitle}{\emph{Proceedings of the 11th Nordic Conference on Human-Computer Interaction: Shaping Experiences, Shaping Society}} (Tallinn, Estonia) \emph{(\bibinfo{series}{NordiCHI '20})}. Article \bibinfo{articleno}{19}, \bibinfo{numpages}{12}~pages.
\newblock
\showISBNx{9781450375795}


\bibitem[superagent({[n.\,d.]})]%
        {super_agent}
superagent \bibinfo{year}{[n.\,d.]}\natexlab{}.
\newblock \bibinfo{title}{superagent – My WordPress Blog}.
\newblock \bibinfo{howpublished}{\url{https://super-agent.com/}}.
\newblock


\bibitem[Tran et~al\mbox{.}(2024)]%
        {Tran-etal-24-CHI}
\bibfield{author}{\bibinfo{person}{Van~Hong Tran}, \bibinfo{person}{Aarushi Mehrotra}, \bibinfo{person}{Marshini Chetty}, \bibinfo{person}{Nick Feamster}, \bibinfo{person}{Jens Frankenreiter}, {and} \bibinfo{person}{Lior Strahilevitz}.} \bibinfo{year}{2024}\natexlab{}.
\newblock \showarticletitle{Measuring Compliance with the California Consumer Privacy Act Over Space and Time}. In \bibinfo{booktitle}{\emph{Proceedings of the CHI Conference on Human Factors in Computing Systems}} (Honolulu, HI, USA) \emph{(\bibinfo{series}{CHI '24})}. Article \bibinfo{articleno}{785}, \bibinfo{numpages}{19}~pages.
\newblock
\showISBNx{9798400703300}


\bibitem[Tranco(2023)]%
        {trancolist}
\bibfield{author}{\bibinfo{person}{Tranco}.} \bibinfo{year}{2023}\natexlab{}.
\newblock \bibinfo{title}{Information on the Tranco list with ID LYLP4}.
\newblock \bibinfo{howpublished}{\url{https://tranco-list.eu/list/LYLP4/}}.
\newblock


\bibitem[Trevisan et~al\mbox{.}(2019)]%
        {4years}
\bibfield{author}{\bibinfo{person}{Martino Trevisan}, \bibinfo{person}{Stefano Traverso}, \bibinfo{person}{Eleonora Bassi}, {and} \bibinfo{person}{Marco Mellia}.} \bibinfo{year}{2019}\natexlab{}.
\newblock \showarticletitle{{4 Years of EU Cookie Law: Results and Lessons Learned}}. In \bibinfo{booktitle}{\emph{Proceedings on Privacy Enhancing Technologies 2019}}.
\newblock


\bibitem[(UODO)(2019)]%
        {UODO-revokeDecision}
\bibfield{author}{\bibinfo{person}{Polish~DPA (UODO)}.} \bibinfo{year}{2019}\natexlab{}.
\newblock \bibinfo{title}{Decision ZSPU.421.3.201 against ClickQuickNow Sp. z o. o.}
\newblock \bibinfo{howpublished}{\url{https://uodo.gov.pl/decyzje/ZSPR.421.7.2019}}.
\newblock


\bibitem[Zengrui~Liu and Saxena(2024)]%
        {Liu-etal-24-popets}
\bibfield{author}{\bibinfo{person}{Umar~Iqbal Zengrui~Liu} {and} \bibinfo{person}{Nitesh Saxena}.} \bibinfo{year}{2024}\natexlab{}.
\newblock \showarticletitle{{Opted Out, Yet Tracked: Are Regulations Enough to Protect Your Privacy?}}
\newblock \bibinfo{journal}{\emph{Proc. Priv. Enhancing Technol.}} \bibinfo{volume}{2024}, \bibinfo{number}{1} (\bibinfo{year}{2024}), \bibinfo{pages}{280--299}.
\newblock
\urldef\tempurl%
\url{https://doi.org/10.56553/popets-2024-0016}
\showDOI{\tempurl}


\bibitem[Zhang et~al\mbox{.}(2024)]%
        {cschecker}
\bibfield{author}{\bibinfo{person}{Mingxue Zhang}, \bibinfo{person}{Wei Meng}, \bibinfo{person}{You Zhou}, {and} \bibinfo{person}{Kui Ren}.} \bibinfo{year}{2024}\natexlab{}.
\newblock \showarticletitle{{CSChecker: Revisiting GDPR and CCPA Compliance of Cookie Banners on the Web}}. In \bibinfo{booktitle}{\emph{Proceedings of the IEEE/ACM 46th International Conference on Software Engineering}} (Lisbon, Portugal). Article \bibinfo{articleno}{174}, \bibinfo{numpages}{12}~pages.
\newblock


\end{thebibliography}

%
\appendix
\section*{Appendix}
\label{sec:appendix}

\balance

\section{Inconsistency between consent returned via {\tcfapi} and consent shared on the network}
\label{sec:anomalies}

\begin{table*}[h]
\begin{tabular}{|p{2.1cm}|p{2.1cm}|p{12.5cm}|}
\hline
\textbf{Responsible party} & \textbf{Website} & \textbf{Reason for inconsistency} \\ \hline                                 
3rd-party (CMP) Ketch Kloud, Inc. & {forbes.com} & \emph{Different \tcs:} number of vendor consents and number of vendor legitimate purposes increased in {\tcs} on the network w.r.t to the {\tcs} returned by the {\tcfapi}.\\ 
\hline
3rd-party (CMP) Ketch Kloud, Inc. & {time.com} & \emph{Different \tcs:} number of vendor consents, number of vendor legitimate purposes and number purpose consent increased  in {\tcs} on the network w.r.t {\tcs} returned by the {\tcfapi}. 13 different {\tcss} received in the responses.
\\ 
\hline
1st-party & {n-tv.de} & \emph{Hardcoded \tcs:} String hardcoded in 1st-party script, sent in the request to 3rd-party. The {\tcs} had additional purpose legitimate interests (5) and vendor legitimate interests (3). \\ 
\hline
3rd-party (CMP) Didomi  & {cadenaser.com} &  Third-party sends a different {\tcs} w.r.t {\tcs} returned by {\tcs} in the response to a request made by a TP which was informed of the correct ({\tcs} returned by \tcfapi) in previous requests. \\ \hline
1st-party & {deadline.com} & Old, non-updated (after acceptance) consent is shared via network requests even after revocation. \\ 
\hline
3rd-party (kinja-static.com)  & {kotaku.com} & Third-party script was loaded after the consent was updated (from the previous stages, both in case of after acceptance and revocation), but it still used the wrong consent.  \\ 
\hline
3rd-party (CMP) InMobi PTE Ltd & {manchester-eveningnews.co.uk}, {walesonline.co.uk} & Although all the requests sent an updated consent after revocation, {signal-beacon.s-onetag.com/beacon.min.js} script continues to send the positive  consent (old consent-after acceptance) \\ \hline
\end{tabular}
\caption{Inconsistencies in consent strings shared over the network} 
\label{tab:anomaly2}
\end{table*}

\begin{table*}
\begin{tabular}{|p{1cm}|p{5.5cm}|p{4.5cm}|p{5cm}|}
\hline
\textbf{Purpose}  & \textbf{Name} & \textbf{Repeated from v2.0} & \textbf{Requires Consent} \\ \hline
1    & Store and/or access information on a device & = 1 & \cmark \\ \hline
2 & Use limited data to select advertising & Looks new but related to no. 2 = “Select basic ads”  & \cmark \\ \hline
3 & Create profiles for personalised advertising & = 3 & \cmark \\ \hline
4 & Use profiles to select personalised advertising & = 4 & \cmark \\ \hline
5 & Create profiles to personalise content & = 5 & (\cmark) Not clear: in principle, this purpose can be legitimized under a legitimate interest, but it would fail the this test. \\ \hline
6 & Use profiles to select personalised content & = 6 & (\cmark) Not clear: in principle, this purpose can be legitimized under a legitimate interest, but it would fail the legitimate interest test. \\ \hline
7 & Measure advertising performance & = 7 & \cmark \\ \hline
8 & Measure content performance & = 8 & \cmark \\ \hline
9 & Understand audiences through statistics or combinations of data from different sources    &New language but looks like the former purpose no. 9, "Apply market research to generate audience insights" & \cmark \\ \hline
10 & Develop and improve services & = 10 & It is not specific, and so we cannot derive its legal basis. \newline X but in principle it could rely on LI though it could fail the LI test \\ \hline
11 & Use limited data to select content & New & X \\ \hline
\end{tabular}
    
\caption{IAB TCF purposes in v2.2 in the {\tcs} and the applicable legal basis. The ``Requires Consent'' column sums up our analysis. The ``Repeated from v2.0.'' column compares the purposes of versions v2.0 and v2.2. We add parentheses if exceptions occur.
You can find the descriptions of all the purposes in the supplementary material~\cite{sup-material}. 
}
\label{tab:purposes}
\end{table*}

Table~\ref{tab:anomaly2} presents different types of  inconsistencies we have observed in our analysis, grouped by the reason for such inconsistency. 
%
%
We observed three types of inconsistencies in the requests and responses: (1) due to delay in updating the consent after user revokes consent; (2) due to introducing/using a different \tcs; (3) due to some scripts not being updated about revocation in consent. 


%
Cases where the consent was not updated immediately are the requests sent to third-parties and containing the {\tcs} before user update. 
We also investigate the time difference between the user update event and when the old consent was still being sent on the network. We additionally observe cases where some URL parameters were updated while some variable in the POST data containing the old consent and taking few seconds to be updated. Examples include \texttt{cnn.com, idnes.cz,independent.co.uk}. 

Cases where a completely different {\tcs} was shared on the network, include two websites \texttt{forbes.com} and \texttt{time.com}, where the CMP's server sends a {\tcs} in the response to a request made by the CMP's script. This {\tcs} is not the same as the one returned by the {\tcfapi} and contains additional vendors and  purposes. We however don't observe this {\tcs} being shared to other third-parties. We observe this behavior both after acceptance and after revocation. In case of \texttt{time.com}, we observed  13 different \tcss. We also observe one case where a completely different {\tcs} with additional legitimate interests in \texttt{n-tv.de} w.r.t the {\tcs} retuned by the {\tcfapi} after both in case of after acceptance and after revocation. In this case, the {\tcs} was hard-coded in a first-party script and sent to a third-party.  In case of \texttt{cadenaser.com}, a third-party server sends a different {\tcs} in response to a third-party script request. However it was not further sent to other third-parties. We also want to point out here that we observe four cases where the ConsentScreen parameter in the {\tcs} is changed. However it does not affect the user's consent in any way, hence is not relevant for the discussion.

Finally, we also observe cases where the scripts do not use the updated consent. While in \texttt{deadline.com}, all the requests use the old consent, in two websites, \texttt{manchestereveningnews.co.uk} and \texttt{walesonline.co.uk} some scripts send requests with the updated consent while the one script (signal-beacon.s-onetag.com/beacon.min.js) still sends the old consent on the network. 
Upon further investigation, we found that this script is used by the CMP but is not the one responsible for updating consent when the user takes action to do so.
%
Table~\ref{tab:anomaly2} also shows that localStorage was not updated while the cookies and {\tcfapi} was. 
It’s possible that the script used the value from local storage instead, resulting in an outdated consent being sent. However, there is no evidence to support this.


\begin{figure*}[b]
  \centering   
  \begin{subfigure}[b]{0.3\textwidth}
    \includegraphics[width=\textwidth]{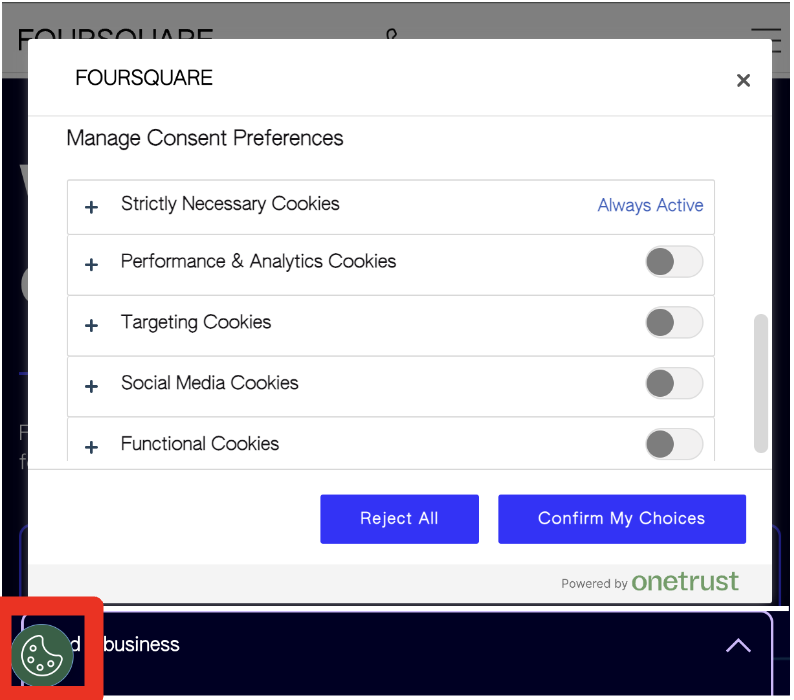}%
    \caption{Persistent icon}
    \label{fig:revocation_explicit1}
  \end{subfigure}
  \hfill
  \begin{subfigure}[b]{0.32\textwidth}
\includegraphics[width=\textwidth]{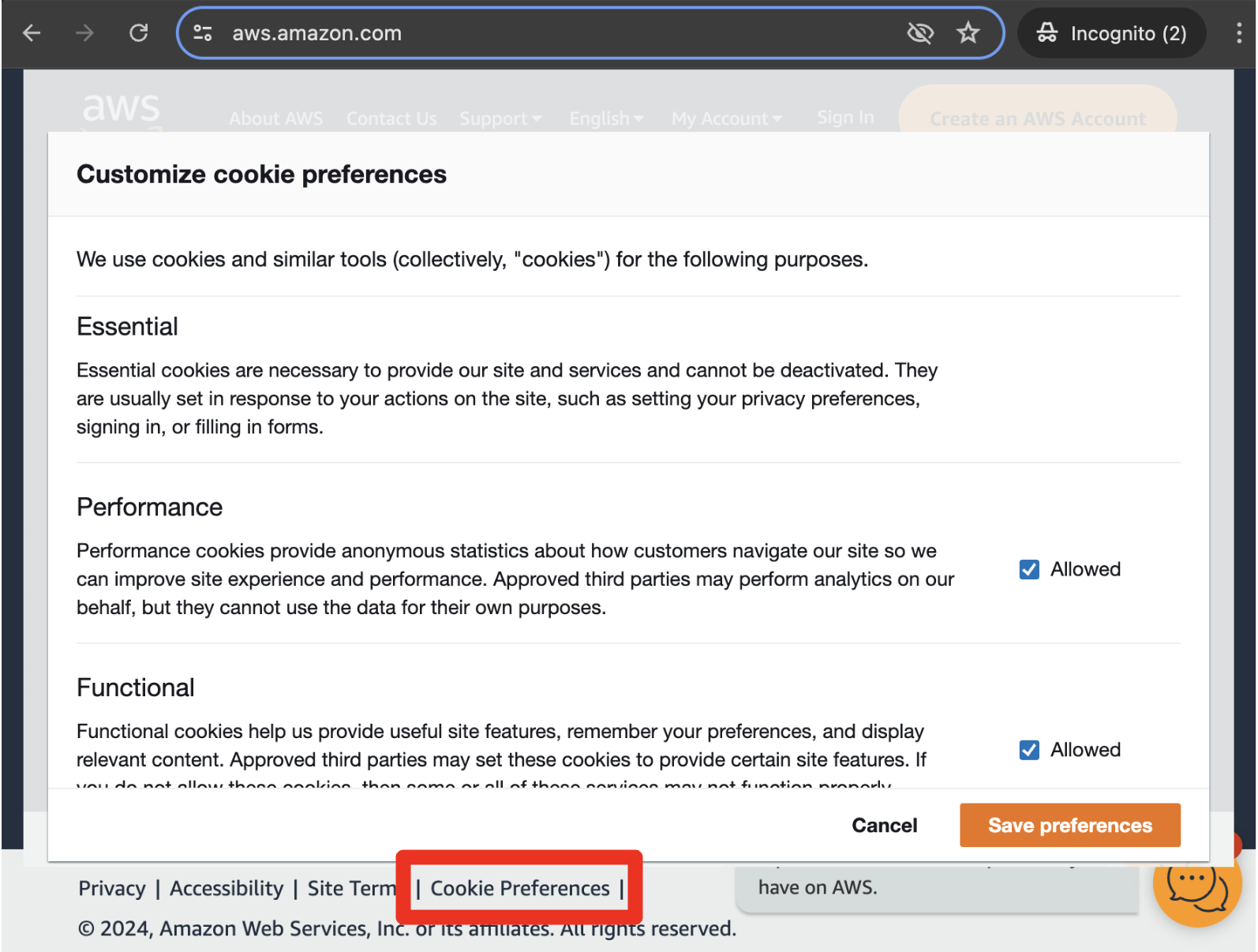}
\caption{Footer options}
\label{fig:revocation_explicit2}
\end{subfigure}
  \hfill
\begin{subfigure}[b]{0.32\textwidth}
  \centering
  \includegraphics[width=\textwidth]{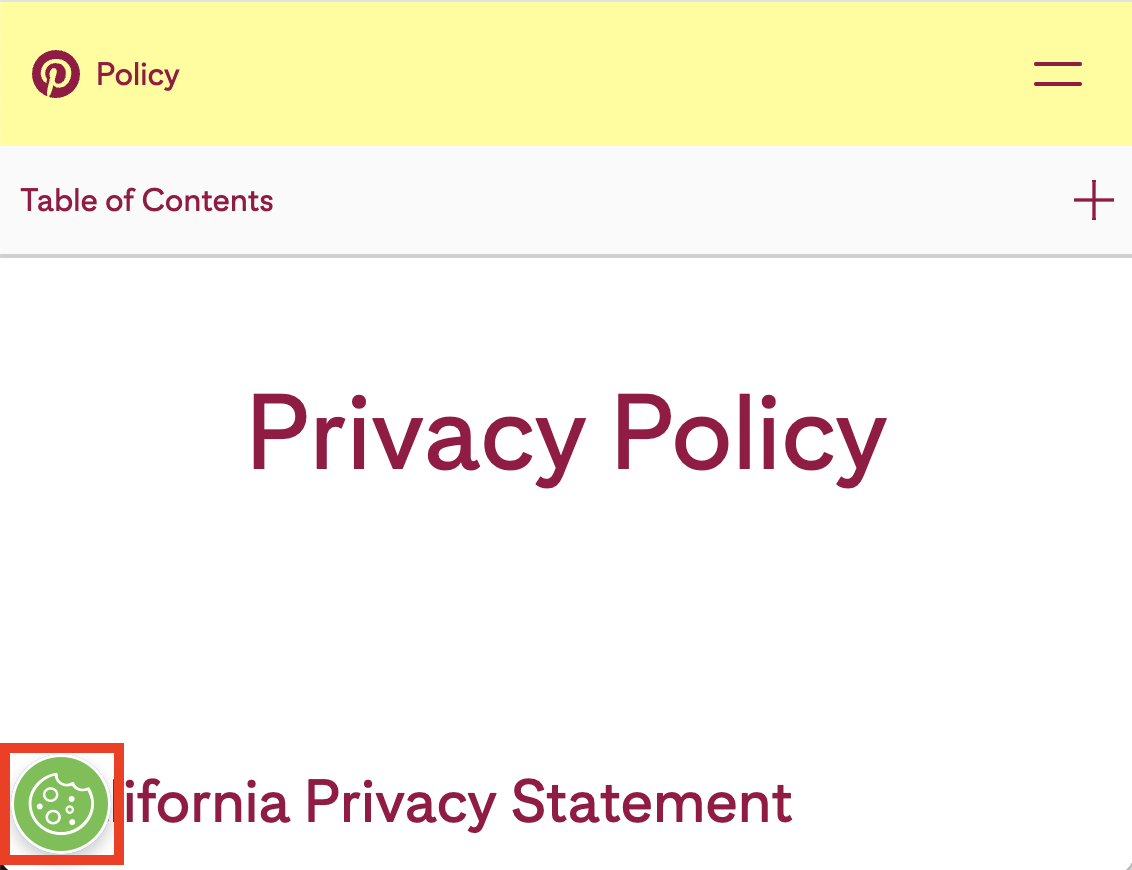}
  \caption{Banner on the policy page}
  \label{fig:revocation_explicit3}
\end{subfigure}
\hfill
\begin{subfigure}[b]{\textwidth}
  \centering
  \includegraphics[width=\textwidth]{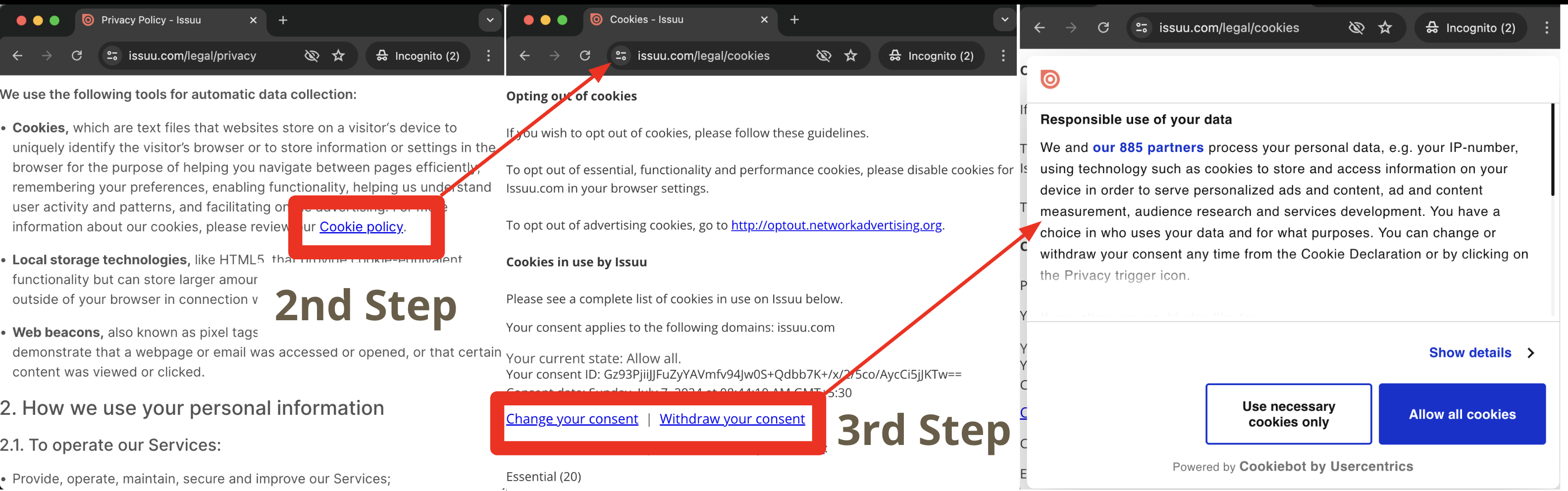}
  \caption{Options on other pages}
  \label{fig:revocation_explicit5}
\end{subfigure}
\caption{Revocation options within the interface on websites \NBtext{Always include URLs of the websites you're showing and the timestamp of the visit in the titles of the screenshots.}}
\label{fig:within_rev}
\end{figure*}

\begin{figure*}[b]
\begin{subfigure}[b]{0.3\textwidth}
  \centering
  \includegraphics[width=\textwidth]{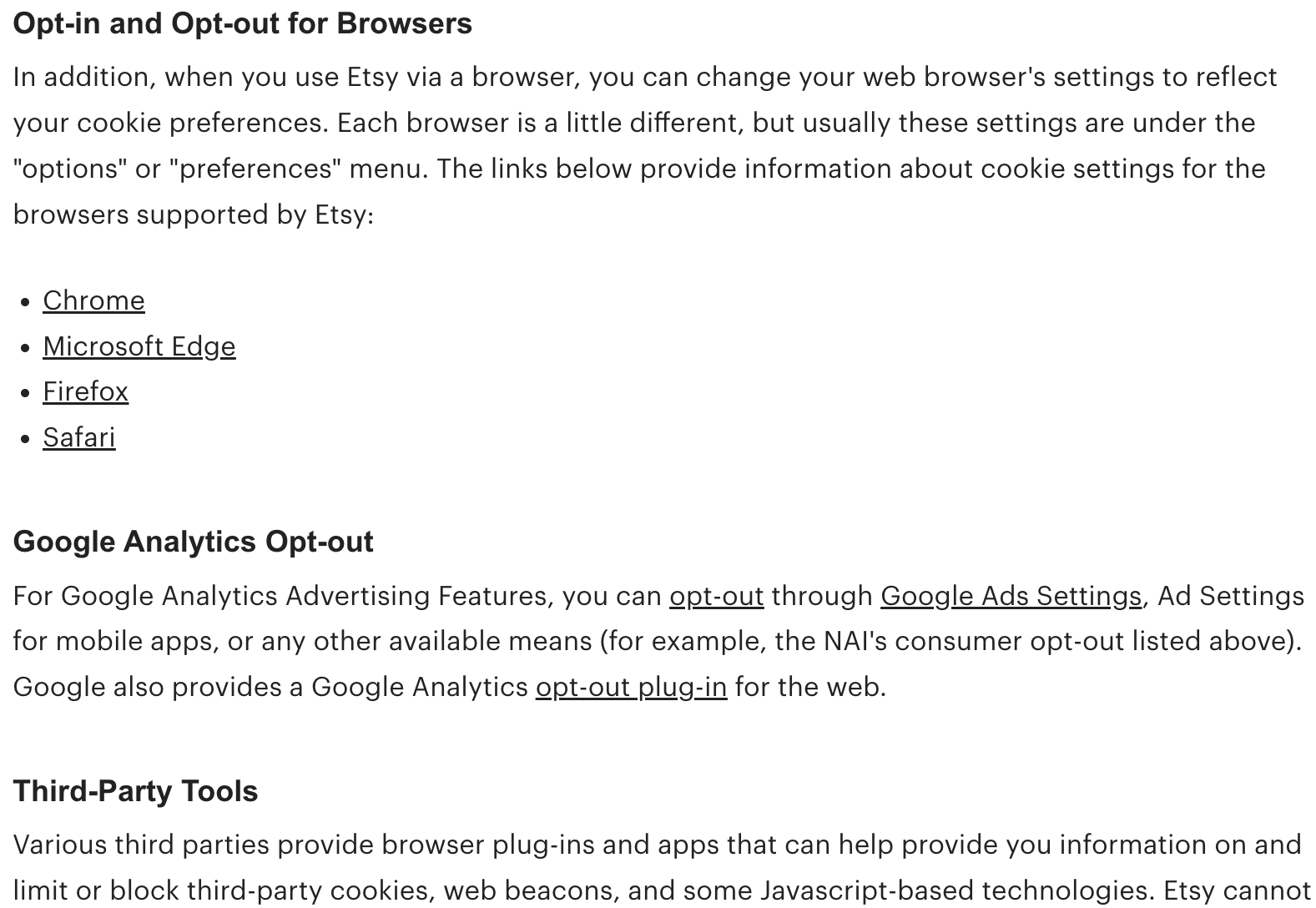}
  \caption{Browser settings or 3rd-party links}
  \label{fig:revocation_suggestive1}
\end{subfigure}
 \hfill
\begin{subfigure}[b]{0.31\textwidth}
  \centering
  \includegraphics[width=\textwidth]{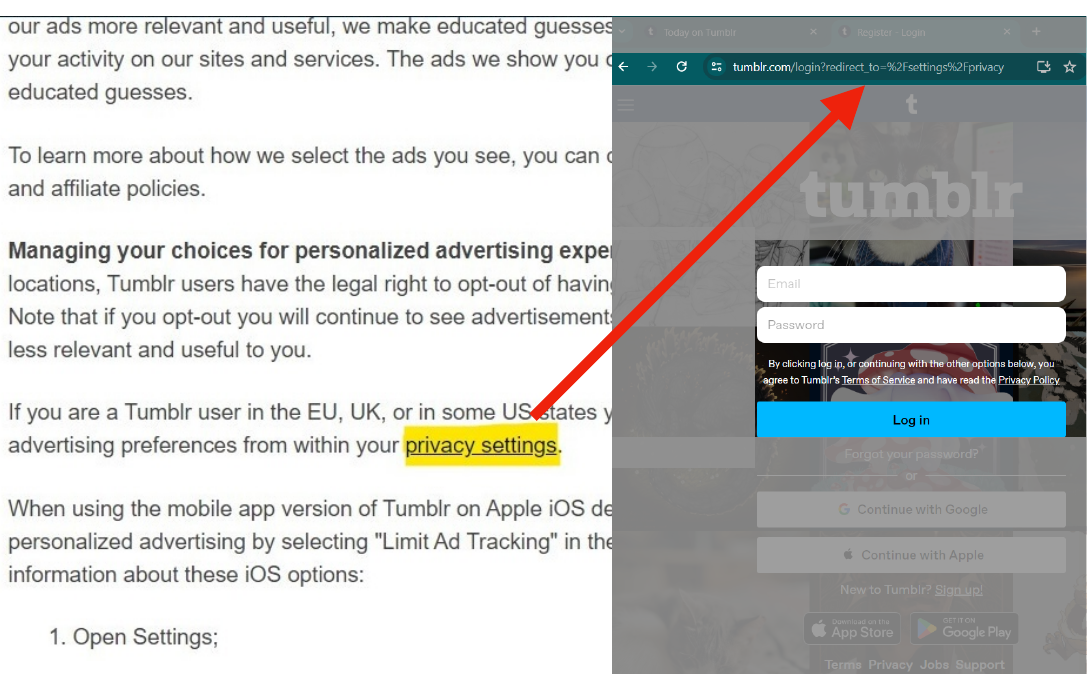}
  \caption{Consent options after login}
  \label{fig:revocation_suggestive2}
\end{subfigure}
 \hfill
\begin{subfigure}[b]{0.32\textwidth}
  \centering
  \includegraphics[width=\textwidth]{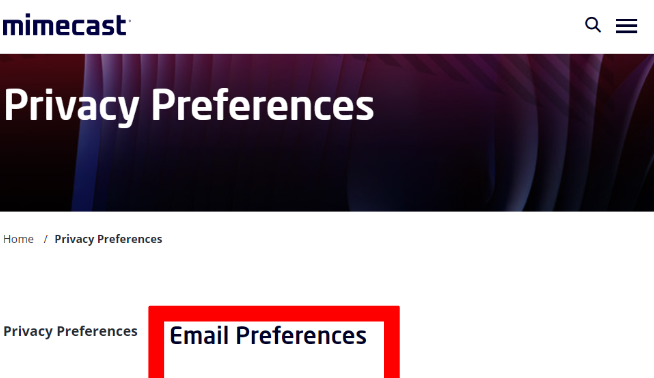}
  \caption{Contact/Email}
  \label{fig:revocation_suggestive3}
\end{subfigure}
 \hfill
\begin{subfigure}[b]{0.3\textwidth}
  \centering
  \includegraphics[width=\textwidth]{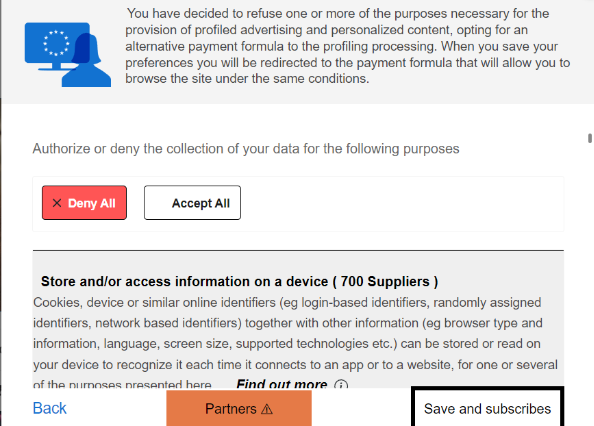}
  \caption{Paywalls}
  \label{fig:revocation_suggestive4}
\end{subfigure}
\hfill
\begin{subfigure}[b]{0.3\textwidth}
  \centering
  \includegraphics[width=\textwidth]{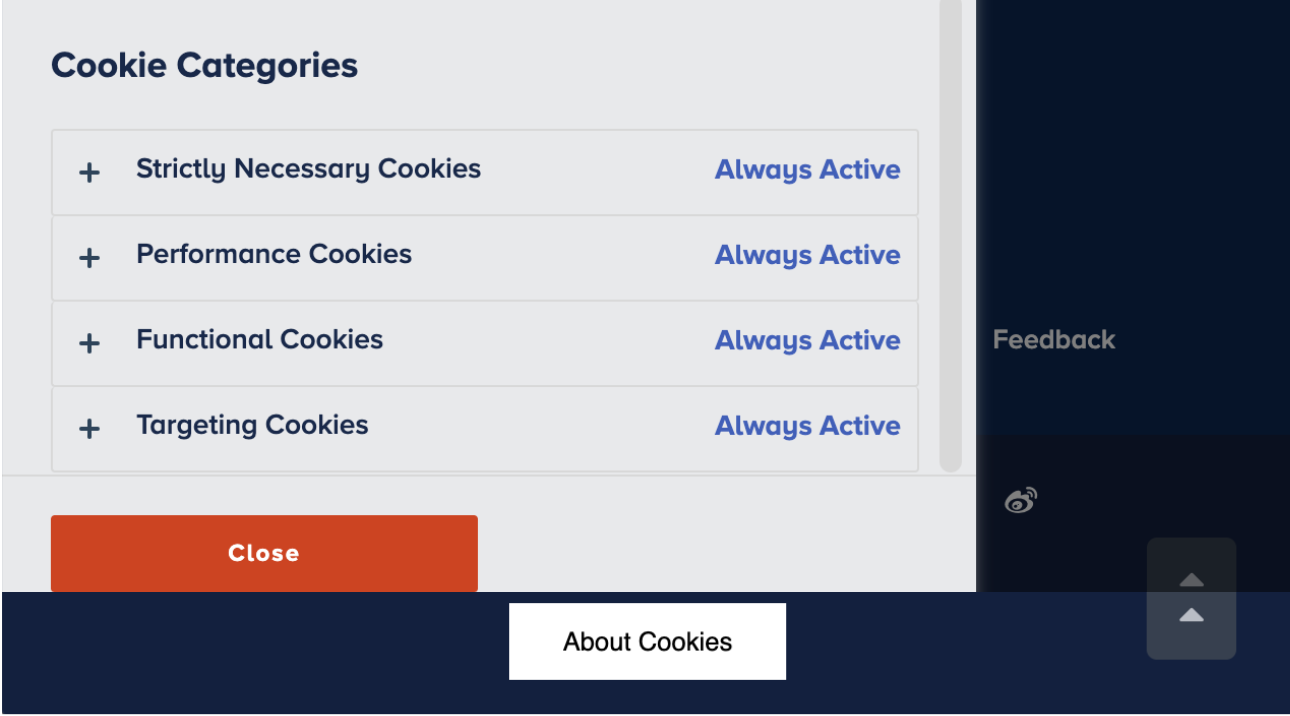}
  \caption{Revocation mentioned but not possible}
  \label{fig:no_revocation1}
\end{subfigure}
\hfill
\begin{subfigure}[b]{0.3\textwidth}
  \centering
  \includegraphics[width=\textwidth]{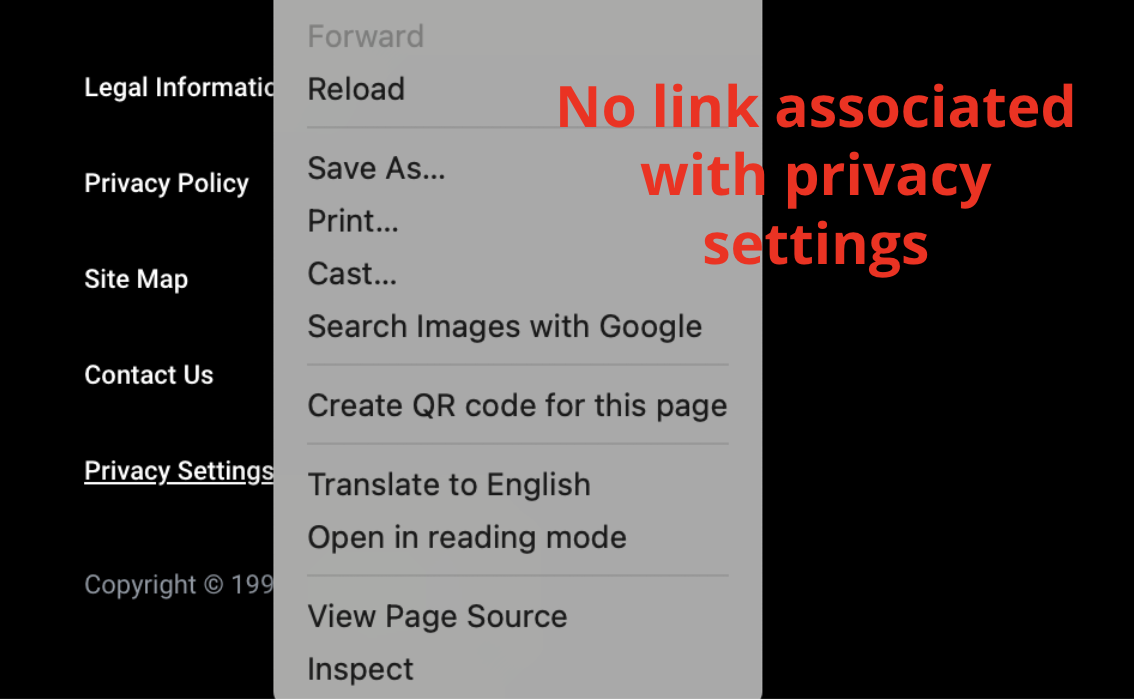}
  \caption{Privacy settings without links}
  \label{fig:no_revocation2}
\end{subfigure}
\caption{Websites with revocation options via different interfaces or no revocation options \NBtext{Always include URLs of the websites you're showing and the timestamp of the visit in the titles of the screenshots.}}
\label{fig:browser_rev}
\end{figure*}

\begin{table*}[ht]
\centering
\begin{tabular}{|p{1cm}|p{1cm}|>{\tgreen}p{1cm}|>{\tgreen}p{1cm}|>{\tgreen}p{1cm}|>{\tyellow}p{1cm}|>{\tred}p{1cm}|>{\tred}p{1cm}|>{\tred}p{1cm}|>{\tred}p{1cm}|>{\tred}p{1cm}|>{\tdred}p{1cm}|}
\hline
\multicolumn{2}{|c|}{} & \multicolumn{4}{c|}{\cellcolor[HTML]{D9EAD3}\textbf{Within the Same Interface}} & \multicolumn{5}{c|}{\cellcolor[HTML]{F4CCCC}\textbf{Via Different Interfaces}} & \cellcolor[HTML]{EA9999}\textbf{No Revocation} \\ \hline
\textbf{Banner} & \textbf{No. of Websites} & \textbf{Icon or Button (0 Steps)} & \textbf{Footer Options (1 Step)} & \textbf{Banner on Policy (1 step)} & \textbf{Options through Policy Page ($\geq$ 2 steps)} & \textbf{Browser set., 3rd-party links} & \textbf{After Login} & \textbf{Contact/ Email} & \textbf{Paywalls} & \textbf{Option mentioned but doesn't work} & \textbf{No revocation} \\ \hline
\textbf{Consent Banner} & 54 & 4 & 27 & 3 & 18 & 2 & 0 & 0 & 0 & 0 & 0 \\ \hline
\textbf{No Option Banner} & 8 & 0 & 0 & 0 & 5 & 2 & 1 & 0 & 0 & 0 & 0 \\ \hline
\textbf{No Banner} & 43 & 0 & 2 & 5 & 7 & 17 & 0 & 3 & 0 & 0 & 4 {[}+5{]} \\ \hline
\textbf{Total} & 105 & 4 & 29 & 8 & 30 & 21 & 1 & 3 & 0 & 0 & 9 \\ \hline
\end{tabular}
\caption{Prevalence and type of consent revocation options on 105 websites (rank 1-200) \emph{without} detected CMPs}
\label{app-tab:data_analysis1}
\end{table*}

\begin{table*}[ht]
\centering
\begin{tabular}{|p{1cm}|p{1cm}|>{\tgreen}p{1cm}|>{\tgreen}p{1cm}|>{\tgreen}p{1cm}|>{\tyellow}p{1cm}|>{\tred}p{1cm}|>{\tred}p{1cm}|>{\tred}p{1cm}|>{\tred}p{1cm}|>{\tred}p{1cm}|>{\tdred}p{1cm}|}
\hline
\multicolumn{2}{|c|}{} & \multicolumn{4}{c|}{\cellcolor[HTML]{D9EAD3}\textbf{Within the Same Interface}} & \multicolumn{5}{c|}{\cellcolor[HTML]{F4CCCC}\textbf{Via Different Interfaces}} & \cellcolor[HTML]{EA9999}\textbf{No Revocation} \\ \hline
\textbf{Banner} & \textbf{No. of Websites} & \textbf{Icon or Button (0 Steps)} & \textbf{Footer Options (1 Step)} & \textbf{Banner on Policy (1 step)} & \textbf{Options through Policy Page ($\geq$ 2 steps)} & \textbf{Browser set., 3rd-party links} & \textbf{After Login} & \textbf{Contact/ Email} & \textbf{Paywalls} & \textbf{Option mentioned but doesn't work} & \textbf{No revocation} \\ \hline
\textbf{Consent Banner} & 54 & 5 & 36 & 3 & 3 & 6 & 1 & 0 & 0 & 0 & 0 \\ \hline
\textbf{No Option Banner} & 0 & 0 & 0 & 0 & 5 & 2 & 1 & 0 & 0 & 0 & 0 \\ \hline
\textbf{No Banner} & 2 & 0 & 0 & 0 & 0 & 0 & 0 & 0 & 0 & 0 & 0 \\ \hline
\textbf{Total} & 56 & 5 & 38 & 3 & 3 & 6 & 1 & 0 & 0 & 0 & 0 \\ \hline
\end{tabular}
\caption{Prevalence and type of consent revocation options on 56 websites (rank 1-200) \emph{with} detected CMPs}   \label{app-tab:data_analysis2}
\end{table*}

\begin{table*}[ht]
\centering
\begin{tabular}{|p{1cm}|p{1cm}|>{\tgreen}p{1cm}|>{\tgreen}p{1cm}|>{\tgreen}p{1cm}|>{\tyellow}p{1cm}|>{\tred}p{1cm}|>{\tred}p{1cm}|>{\tred}p{1cm}|>{\tred}p{1cm}|>{\tred}p{1cm}|>{\tdred}p{1cm}|}
\hline
\multicolumn{2}{|c|}{} & \multicolumn{4}{c|}{\cellcolor[HTML]{D9EAD3}\textbf{Within the Same Interface}} & \multicolumn{5}{c|}{\cellcolor[HTML]{F4CCCC}\textbf{Via Different Interfaces}} & \cellcolor[HTML]{EA9999}\textbf{No Revocation} \\ \hline
\textbf{Banner} & \textbf{No. of Websites} & \textbf{Icon or Button (0 Steps)} & \textbf{Footer Options (1 Step)} & \textbf{Banner on Policy (1 step)} & \textbf{Options through Policy Page ($\geq$ 2 steps)} & \textbf{Browser set., 3rd-party links} & \textbf{After Login} & \textbf{Contact/ Email} & \textbf{Paywalls} & \textbf{Option mentioned but doesn't work} & \textbf{No revocation} \\ \hline
\textbf{Consent Banner} & 214 & 33 & 125 & 0 & 26 & 9 & 0 & 3 & 11 & 3 & 7 \\ \hline
\textbf{No Option Banner} & 0 & 0 & 0 & 0 & 0 & 0 & 0 & 0 & 0 & 0 & 0 \\ \hline
\textbf{No Banner} & 11 & 0 & 8 & 0 & 0 & 1 & 0 & 0 & 0 & 0 & 2 \\ \hline
\textbf{Total} & 225 & 33 & 133 & 0 & 26 & 10 & 0 & 3 & 11 & 3 & 9 \\ \hline
\end{tabular}
\caption{Prevalence and type of consent revocation options on 200 to 5000 ranked websites \emph{with} detected CMPs}
  \label{tab:data_analysis3}
\end{table*}

\end{document}